\documentclass[traditabstract]{aa} 
\usepackage[dvips]{color}
\usepackage{graphicx}
\usepackage{txfonts}
\usepackage{natbib}

\begin{document}
\bibliographystyle{aa}
   \title{General relativistic radiative transfer: formulation and emission from structured tori around black holes}

   \authorrunning{Younsi, Wu and Fuerst}
  \titlerunning{General relativistic radiative transfer}

   \author{Ziri Younsi,
          \inst{1}\fnmsep\thanks{Send offprint requests to: Z.~Younsi}
          Kinwah Wu\inst{1}
          \and 
          Steven V.~Fuerst\inst{2}
          }

   \institute{Mullard Space Science Laboratory, University College London,
                   Holmbury St Mary, Surrey, RH5 6NT, UK \\
              \email{zy2@mssl.ucl.ac.uk, kw@mssl.ucl.ac.uk}           
         \and
             Kavli Institute for Particle Astrophysics and Cosmology,
             Stanford University, Stanford, CA 94305, USA   }

   \date{Received March XX, 2012}

 
\abstract{}{}{}{}{}
\abstract 
{We aim to construct a general relativistic radiative transfer formulation, applicable to particles with or without mass in astrophysical settings, 
   wherein ray-tracing calculations can be performed for arbitrary geodesics for a given space-time geometry.  
The relativistic radiative transfer formulation is derived from first principles: conserving particle number and phase-space density.   
The formulation is covariant, 
   and transfer calculations are conducted along particle geodesics 
   connecting the emitters and the observer. 
The geodesics are determined through the space-time metric, which is specified beforehand.    
Absorption and emission in the radiative transfer calculations are treated explicitly.  
The particle-medium interaction is evaluated in the local inertial frame, co-moving with the medium.   
Relativistic, geometrical and optical depth effects are treated self-consistently within 
  an integral covariant framework. 
We present a self-consistent general relativistic radiative transfer formulation 
  with  explicit treatment of emission and absorption. 
The formulation is general and is applicable to both particles with mass and without mass.    
The presence of particles has two major effects:  
 firstly the particle bundle ray is no longer along the null geodesic, 
 and secondly the intensity variation along the particle bundle ray 
 is reduced by an aberration factor.   
The radiative transfer formulation 
  can handle 3D geometrical settings  
  and structured objects with variations and gradients in the optical depths across the objects and along the line-of-sight.    
Such scenarios are applicable in calculations of photon emission from complex structured accretion flows around black holes 
  and neutrino emission from remnant neutron tori in neutron-star mergers.  
We apply the formulation and demonstrate radiation transfer calculations 
  for emission from accretion tori around rotating black holes. 
We consider two cases: 
  idealised optically thick tori that have a sharply defined emission boundary surface, 
  and structured tori that allow variations in the absorption coefficient and emissivity within the tori. 
We show intensity images and emission spectra of the tori obtained in our calculations. 
Our findings in the radiative transfer calculations are summarised as follows. 
(i) Geometrical effects, such as lensing-induced self-occulation and multiple-image contribution  
     are much more significant in accretion tori than geometrically thin accretion disks. 
(ii) Optically thin accretion tori show emission line profiles 
     distinguishable from the profiles of lines from optically thick accretion tori and lines from optically thick geometrically thin accretion disks. 
(iii) The line profiles of the optically thin accretion tori have a
    weaker dependence on the viewing inclination angle than those of the optically thick accretion tori or accretion disks, 
    especially at high viewing inclination angles. 
(iv) Limb effects are present in accretion tori with finite optical depths, 
    due to density and temperature stratification within the tori.
We  note that  in accretion flows onto relativistic compact objects, gravitationally induced line resonance can occur. 
This resonance occurs easily in 3D flows, but not in 2D flows, such as a thin accretion disk around a black hole.}    

\keywords{accretion, accretion disks -- black hole physics -- galaxies: active -- line: profiles -- radiative transfer -- relativity}

\maketitle
%

\section{Introduction}

   The X-ray emission observed in active galactic nuclei (AGN) and black
      hole binaries is believed to be powered by accretion of material 
      onto black holes  \citep{Salpeter1964, Lynden-Bell1969, Shakura1973}.  
   The accreting hot plasmas rotating around the black hole form  a disk or a  torus. 
   It has been suggested that dense remnant neutron tori can also be formed 
     around compact objects, which could be a very massive neutron star or a black hole, 
     after two neutron stars merge \citep[see][]{Shibata2003, Baiotti2008, Rezzolla2010}.        
   In such systems, the influence of curved space-time is significant. 
   It affects the radiative transport of particles in the accretion flow  
      as well as the hydrodynamics of the flow itself \citep[see][]{Novikov1973}.  
       
   Emission from accretion disks around compact objects has been investigated for several decades now. 
   Emission lines from geometrically thin accretion disks around gravitating objects 
      are expected to have two peaks \citep{Smak1969}.  
   The peaks correspond to emission from the two parts of the disk, which have opposite projected line-of-sight velocities. 
   Double-peaked optical lines have been observed in a variety of binary systems 
       \citep[e.g.\ black hole X-ray binaries,][]{Johnston1989, Marsh1994, Soria1999, Wu2001}.    
   Double-peaked optical lines are also seen in a small fraction of AGN  
      \citep{Puchnarewicz1996, Eracleous2003, Strateva2006}.  
   These double-peaked lines can be explained in a Newtonian framework as described in \cite{Smak1969} 
      \citep[see also][]{Horne1986}.  
   Double-peaked lines have also been observed in the X-ray spectra of accreting black holes. 
   Broad asymmetric double-peaked Fe K$\alpha$ lines were found in the spectra of a number of AGN \citep[e.g.\ MCG -6-30-15,][]{Tanaka1995}. 
   The X-rays of AGN are believed to originate from regions very close to the central black hole,  
      where the accretion flow is highly relativistic and the gravity is strong. 
   The emissions from different parts of the accretion flow are therefore boosted differentially.  
   Various relativistic effects also cause additional differential broadening and distortion of any line emission.  
   As such, the emission lines from the inner regions of relativistic accretion disks around black holes 
     have a very broad profile,  
     with an extended red wing and a pronounced blue peak \citep{Cunningham1975, Reynolds1999, Fabian2000}.  
   Black holes with faster spins would give rise to relativistic accretion lines with a broader red wing, 
    because the inner boundary of an accretion disk around a maximally rotating black hole can extend very close to the black hole event horizon.  
   
%

The emission from accreting gas and outflows in the vicinity of black holes is subject to Doppler shifts, lensing, gravitational time-dilation, and other relativistic dynamical effects. 
   There have been numerous calculations of relativistic (photon) lines from accretion disks and tori 
   around black holes 
   \citep[e.g.][]{Cunningham1975, Gerbal1981, Fabian1989, Stella1990, Kojima1991, Bao1992, Fanton1997, 
    Reynolds1999, Fabian2000, 
    Fuerst2004, Beckwith1_2004,  Beckwith2005, Cadez2006, Schnittman2006, Fuerst2007, Dexter2009, Sochora2011, Vincent2011, Wang2012}.  
   The three most common methods for calculating relativistic line profiles  
   are (i) the transfer function method \citep[e.g.][]{Cunningham1975, Fabian2000}, 
   (ii) the elliptic function method \citep[e.g.][]{Dexter2009} and (iii) the direct geodesic
   integration method \citep[e.g.][]{Fuerst2004, Schnittman2006, Fuerst2007, Anderson2010, Vincent2011}.  
   The transfer function method and the elliptic function method are efficient  
      for the calculation of emission from thin axisymmetric optically thick accretion disks 
      but  are not applicable to systems that lack the appropriate geometry and symmetry. 
   The direct geodesic integration method is a brute force approach,  
     and less restrictive in this context, compared to the other two methods. 
   It works well with any three-dimensional (3D) accretion flow, 
      e.g. time-dependent accretion flows from numerical relativistic hydrodynamic simulations, 
      and can also handle opacity variations within the system.   
  In most of these relativistic calculations, the focus was on the investigation of line broadening 
      due to relativistic effects. 
While relativistic ray-tracing of photons in strong gravity 
  and the corresponding calculations of emission line profile broadening 
  have been investigated in various astrophysical settings for decades,  
  there have been only a few studies, often in restricted settings, of the opacity effects 
   due to self-absorption, emission and scattering within the accretion flows 
   and along the line of sight \citep[e.g.][]{Zane1996, Fuerst2006, Wu2006, Wu2008, Dolence2009}.  
Covariant radiative transfer calculations of emission from accretion flows in more general settings 
  with an explicit treatment of absorption, emission and scattering are lacking. 
There is also no corresponding covariant radiative transport formulation applicable 
  to both relativistic particles without mass (photons) and with mass (e.g.\ neutrinos), 
  in a general astrophysical setting in the present literature.    
(Note that particles with mass do not follow null geodesics, 
  and the relativistic formulation and the corresponding ray-tracing need to be modified.) 

   In this work we present the general covariant formulation for radiative transport of relativistic particles.  
   The radiative transfer equation is derived from the Lorentz-invariant form of the conservation law.  
   It explicitly includes emission and absorption processes. 
  The formulation is not restricted to general relativistic radiative transfer of photons.  
  It is general and can be applied to both relativistic particles without mass (e.g.\ photons) 
     and with mass (e.g.\ relativistic electrons and neutrinos).  
  The radiative transfer recovers its conventional form in the Newtonian limit. 
  We carried out demonstrative radiative transfer calculations  
     of emission from model accretion tori around rotating black holes with different optical thicknesses.   
  Our calculations show the convolution of geometrical and optical depth effects, 
    such as limb darkening and brightening caused by optical-depth and emissivity variations, 
      multiple image contribution, and absorption and self-occultation induced by gravitational lensing, which are
    characteristic of 3D structured flows in strong gravity environments. 
   The paper is organised as follows. 
   In Sec. 2 the covariant radiative transfer equation is derived and its solution discussed. 
   In Sec. 3 we show the construction of two torus models, 
      one with a specific sharp emission boundary surface, 
      and another with a stratified density and temperature structure. 
    In Sec. 4 we present the results of radiative transfer calculations for the torus models. 
    We briefly discuss the astrophysical implications of our calculations 
       and the possible occurrence of gravitationally induced line resonance in 3D accretion flows near black holes. 
   

\section{Covariant radiative transfer formulation}

The emission and absorption processes may be considered as sources and sinks in the medium  
   through which the ray bundle of particles is transported. 
In their absence, the number of particles along a ray bundle is conserved.      
In the presence of emission and absorption,  the radiative transfer equation can be expressed as  
\begin{equation}
\frac{dI_{\nu}}{ds}=-\alpha_{\nu}I_{\nu}+j_{\nu} \ , 
\label{rte}
\end{equation}
  where $I_{\nu}\equiv I_{\nu}(s)$ is the specific intensity of the ray at a frequency $\nu$, 
  and $\alpha_{\nu}$ and $j_{\nu}$ are, respectively, the absorption and emission coefficients at a frequency $\nu$. 
By introducing the variable   
\begin{equation}
  \tau_{\nu}(s)=\int_{s_{0}}^{s}\alpha_{\nu}(s')\ ds'   \ , 
\end{equation} 
  which is the optical depth (optical thickness of the medium between $s$ and $s_{0}$),  
  we may rewrite the radiative transfer equation as 
\begin{equation}
\frac{dI_{\nu}}{d\tau_{\nu}}=-I_{\nu}+\frac{j_{\nu}}{\alpha_{\nu}} = -I_{\nu}+S_{\nu}  \  , 
\label{rte2}
\end{equation} 
   where $S_{\nu} = j_{\nu}/\alpha_{\nu}$ is the source function.  
Direct integration of the equation yields       
\begin{eqnarray} 
   I_{\nu}(s)  & = &   I_{\nu}(s_{0})\ {\rm e}^{-\tau_{\nu}}+\int_{s_{0}}^{s} j_{\nu}(s') \ {\rm e}^{-(\tau_{\nu}(s)-\tau_{\nu}(s'))} ds'  \nonumber  \\ 
  & = &  I_{\nu}(0)\ {\rm e}^{-\tau_{\nu}} + \int_{0}^{\tau_{\nu}}S_{\nu}(\tau_{\nu}')\ {\rm e}^{-(\tau_{\nu}-\tau_{\nu}')} d\tau_{\nu}'      \  ,   
\end{eqnarray}  
   where the constant $I_{\nu}(s_{0})$ ($=I_{\nu}(0)$) is the initial value of the specific intensity.   
While optical depth, which is a scalar quantity, is invariant under Lorentz transformations, 
   the radiative transfer equations in the conventional form (equations [\ref{rte}] and [\ref{rte2}]) are not.

We now show that a covariant formulation for radiative transfer can be derived 
  from the conservation of phase space volume and the conservation of particle number.   
Construct a phase space volume $\mathcal{V}$ threaded with a small bundle of particles.  
In the co-moving frame, these particles occupy a spatial volume element $d^{3}{\vec x}=dx \ \! dy \ \! dz$  
   and a momentum volume element  $d^{3}\vec{p}=dp_{x}\ \! dp_{y} \ \! dp_{z}$.  
Liouville's theorem ensures that the phase space volume, given by $d\mathcal{V} =d^{3}{\vec x}d^{3}{\vec p}$,   
   is unchanged along the affine parameter $\lambda$, i.e.\      
\begin{equation}
\frac{d\mathcal{V}}{d\lambda} = 0  \    
\end{equation}
   \citep{MTW1973}. 
This together with the conservation of the number of particles $dN$ 
   within the phase space volume element implies   
   that the phase space density 
\begin{equation} 
   f(x^{i},p^{i}) = \frac{dN}{d\mathcal{V}}   \  ,
\end{equation} 
   is invariant along $\lambda$.  

For relativistic particles, $|\vec{p}|=E$, and $d^{3}\vec{p}=E^{2}dE \ \! d\Omega$ (here and hereafter, unless otherwise stated, we adopt the convention that the speed of light in vacuum $c =1$).    
The volume element of a bundle of relativistic particles is also given by $d^{3}{\vec x}=dA \ dt$, where $dA$ is the area element of the bundle. 
Thus, we may express the phase space density of a bundle of relativistic particles as 
\begin{equation}
  f(x^{i},p^{i})=\frac{dN}{E^{2}dA\ \! dt \ \! dE \ \! d\Omega}\ .
\end{equation}
The specific intensity of a ray (bundle of photons) is simply  
\begin{equation}
  I_{E}=\frac{E \ \! dN}{dA\ \! dt \ \! dE \ \! d\Omega} \   
\end{equation}  
   \citep[see][]{Rybicki_Lightman}.  
It follows that   
\begin{equation} 
  \mathcal{I}  \equiv \frac{I_{\nu}}{\nu^{3}} =  \frac{I_{E}}{E^{3}}  \   
\end{equation}  
  is a Lorentz invariant quantity. 
We denote $ \mathcal{I}$ as the Lorentz-invariant intensity, 
  and it can also be regarded as the occupation number of particles in the phase space for a particle bundle. 
We may also obtain the corresponding Lorentz-invariant absorption coefficient  $\chi = \nu \ \! \alpha_{\nu}$ 
 and the Lorentz-invariant emission coefficient $\eta = j_{\nu}/\nu^{2}$.  
These two coefficients, as seen by the observer, are related to their counterparts in the local rest frame of the medium 
   via $\nu \ \!  \alpha_{\nu} = \nu_{0} \ \! \alpha_{0,\nu}$ and $j_{\nu}/\nu^{2} = j_{0,\nu}/\nu_{0}^{2}$ respectively, 
   where the subscript $``0"$ denotes variables measured in the local rest frame.  
 
\subsection{Photon and relativistic massless particle} 

For photons (or a massless relativistic particle) $k_{\alpha}k^{\alpha}=0$, 
  where $k_{\alpha}$ is the (covariant) 4-momentum.   
Consider a photon propagating in a fluid with 4-velocity $u^{\beta}$.     
 The photon's velocity in the co-moving frame of fluid, $v^{\beta}$, 
  can be obtained by projecting the photon's 4-momentum into the fluid frame, i.e.\ 
\begin{eqnarray}
  v^{\beta}&=&P^{\alpha \beta}k_{\alpha} \nonumber \\
 &=& k^{\beta}+(k_{\alpha}u^{\alpha})u^{\beta} \  , 
\end{eqnarray} 
  where $P^{\alpha \beta}=g^{\alpha \beta}+u^{\alpha}u^{\beta}$ 
  is the projection tensor and $g^{\alpha \beta}$ the space-time metric tensor.     
The variation in the path length $s$ with respect to the affine parameter $\lambda$ is then 
\begin{eqnarray}
\frac{ds}{d\lambda}&=& -||v^{\beta}|| \Big{|}_{\lambda_{obs}} \nonumber \\
                                  &=& -\sqrt{g_{\alpha \beta}(k^{\alpha}+(k_{\beta}u^{\beta})u^{\alpha})(k^{\beta}+(k_{\alpha}u^{\alpha})u^{\beta})}\ \! \Big{|}_{\lambda_{obs}} \nonumber \\
                                  &=& -\sqrt{k_{\beta}k^{\beta}+(k_{\alpha}u^{\alpha})^{2}u_{\beta}u^{\beta}+2(k_{\alpha}u^{\alpha})^{2}}\ \! \Big{|}_{\lambda_{obs}}\nonumber \\
                                  &=& -k_{\alpha}u^{\alpha}\ \! \Big{|}_{\lambda_{obs}}. 
\label{ds}                                
\end{eqnarray}
Here we have used $[-,+,+,+]$ convention for the signature of the space-time metric.     
For a stationary observer located at infinity, $p_{\beta}u^{\beta}=-E_{obs}$.  
The relative energy shift of the photon between the observer's frame and the comoving frame is therefore 
\begin{equation}
   \gamma^{-1} = \frac{\nu_{0}}{\nu}
   =\frac{-k_{\alpha}u^{\alpha}|_{\lambda}}{E_{obs}}
   =\frac{k_{\alpha}u^{\alpha}|_{\lambda}}{k_{\beta}u^{\beta}|_{\lambda_{obs}}}.
\end{equation}

Making use of the Lorentz-invariant properties of the variables $\mathcal{I}$, $\chi$ and $\eta$, and of the optical depth $\tau_{\nu}$, 
  we may rewrite the radiative transfer equation in the following form 
\begin{equation} 
  \frac{d\mathcal{I}}{d\tau_{\nu}}=- \mathcal{I} +\frac{\eta}{\chi}  = - \mathcal{I} +\mathcal{S}  \   , 
\label{cov0}
\end{equation}   
  where $\mathcal{S}=\eta/\chi$ is the Lorentz-invariant source function.  
All quantities in equation (\ref{cov0}) are Lorentz invariant, and hence the equation is covariant.  
As $d\tau_{\nu} = \alpha_{\nu} ds$, we have     
\begin{equation} 
  \frac{d\mathcal{I}}{ds}= - \alpha_{\nu} \mathcal{I} +\frac{j_{\nu}}{\nu^{3} }  \  .     
\label{cov-rte2}
\end{equation}    
It follows that 
\begin{equation}
  \frac{d\mathcal{I}}{d\lambda}= -k_{\alpha}u^{\alpha}\vert_{\lambda}\left(-\alpha_{0,\nu}\mathcal{I}+\frac{j_{0,\nu}}{\nu^{3}} \right) \ , 
\label{cov-rte}
\end{equation} 
  where $\alpha_{0,\nu}\equiv \alpha_{0}(x^{\beta},\nu)$ and $j_{0,\nu}\equiv j_{0}(x^{\beta},\nu)$, where, as before,
   ``0'' denotes variables that are evaluated in the local rest frame, for a given $\nu$. In ray-tracing calculations we specify an observer frequency $\nu$, determine all required variables at this location, and find how these variables change in different reference frames through the radiative transfer equation.
The solution to equation (\ref{cov-rte}) is 
\begin{eqnarray}
  \mathcal{I}(\lambda)&=&\mathcal{I}(\lambda_{0})\mathrm{e}^{-\tau_{\nu}(\lambda)} \nonumber \\
  & & \hspace*{-1.0cm}-\int_{\lambda_{0}}^{\lambda}\frac{j_{0,\nu}(\lambda'')}{\nu^{3}}
\exp \bigg(- \int_{\lambda''}^{\lambda}\alpha_{0,\nu}(\lambda')k_{\alpha}u^{\alpha}|_{\lambda'}d\lambda' \bigg)k_{\alpha}u^{\alpha}|_{\lambda''}d\lambda''  
\label{solution}
\end{eqnarray}
  \citep[cf.][]{Baschek1997,Fuerst2004}, 
  where the optical depth is 
\begin{equation}
\tau_{\nu}(\lambda)=- \int_{\lambda_{0}}^{\lambda}\alpha_{0,\nu}(\lambda^{'})k_{\alpha}u^{\alpha}\vert_{\lambda'}d\lambda'  \ .
\end{equation} 
In terms of the optical depth, 
\begin{equation}
  \mathcal{I}(\tau_{\nu})=\mathcal{I}(\tau_{0})\mathrm{e}^{-\tau_{\nu}}
 + \int_{\tau_{0}}^{\tau_{\nu}}\mathcal{S}(\tau_{\nu}') \mathrm{e}^{-(\tau_{\nu}-\tau_{\nu}')}d\tau_{\nu}' \ .
\end{equation}

For a distant observer, $-k^{\alpha}u_{\alpha}\vert_{\lambda_{obs}}\rightarrow E$, the observed photon energy, which may be normalised to unity.   
The radiative transfer equation can be expressed as two decoupled differential equations
   \begin{eqnarray}
   \frac{d\tau_{\nu}}{d\lambda} &=& \gamma^{-1}\alpha_{0,\nu}  \ ,  \\
   \frac{d\mathcal{I}}{d\lambda} &=& \gamma^{-1}\bigg(\frac{j_{0,\nu}}{\nu^{3}} \bigg)\mathrm{e}^{-\tau_{\nu}}  \  .
   \end{eqnarray}  
These two equations are more useful in practical relativistic radiative transfer calculations because they allow the efficient computation, through a simple Eulerian method, of the optical depth along a ray, regardless of whether the ray-tracing is executed forward or backwards in time. This is then used to compute the intensity along the ray. 
  
\subsection{Particles with mass}   

For massless particles, contraction of the 4-momentum gives $k_{\alpha}k^{\alpha}=0$, 
  but for particles with a non-zero mass $m$, it gives $p_{\alpha}p^{\alpha}=-m^{2}$. 
The presence of the particle mass modifies the particle's equations of motion, changing the geodesics from null to time-like 
  \citep{Carter1968,BoyerLindquist1967}.   
Moreover, a covariant particle flux is mass dependent.   
The radiative transfer equations derived for massless particles (equations (\ref{cov-rte2}) and (\ref{cov-rte}))  
 are therefore not applicable for particles with mass, such as neutrinos or relativistic electrons. 
Nevertheless, the formulation obtained for massless particles can easily be modified to include the effects due to the particle mass. 
We follow a similar procedure as in the case of massless particles in our derivation.   
We express  the variation in the path length with respect to the affine parameter as 
\begin{eqnarray}
  \frac{ds}{d\lambda}                                  
  &=& -\sqrt{g_{\alpha \beta}(p^{\alpha}+(p_{\beta}u^{\beta})u^{\alpha})(p^{\beta}+(p_{\alpha}u^{\alpha})u^{\beta})}\ \! \Big{|}_{\lambda_{obs}} \  ,          
\label{ds2}                                
\end{eqnarray}   
  analogous to equation (\ref{ds}) in the case of massless particles. 
Because $p^{\alpha}p_{\alpha} \neq 0$, we have 
\begin{eqnarray}
  \frac{ds}{d\lambda} &=& 
          -\sqrt{p_{\beta}p^{\beta}+(p_{\alpha}u^{\alpha})^{2}}\  \bigg\vert_{\lambda_{obs}} \nonumber \\
                                  &=& -\sqrt{(p_{\alpha}u^{\alpha})^{2}-m^{2}}\  \bigg\vert_{\lambda_{obs}} \ . 
\end{eqnarray}  
Insert this into equation (\ref{cov0}), which does not depend on the particle mass explicitly. 
After some algebra, we obtain the general covariant transfer equation for relativistic particles 
\begin{equation}
  \frac{d\mathcal{I}}{d\mathcal{\lambda}} 
  = -\sqrt{1-\left(\frac{m}{\ \ p_{\beta}u^{\beta}\big\vert_{\lambda_{obs}}}\right)^{2}}p_{\alpha}u^{\alpha} \bigg\vert_{\lambda} 
  \left(-\alpha_{0,\nu}\mathcal{I}+\frac{j_{0,\nu}}{\nu^{3}}\right) \ . 
\label{mass-rte} 
\end{equation}  

For a stationary observer located at infinity, $p_{\beta}u^{\beta}=-E$. 
Equation (\ref{mass-rte}) differs from equation (\ref{cov-rte}) by an aberration factor   $\sqrt{1-(m/E)^{2}}$.  
This factor reduces the intensity gradient along the ray.  
Because it approaches unity when $m\rightarrow 0$,    
   the radiative transfer equation (\ref{cov-rte}) is the radiative transfer equation (\ref{mass-rte}) in the zero-mass limit.

\section{Accretion tori around black holes} 

We can see in equation (\ref{mass-rte}) that 
  the essential components of the covariant radiative transfer formulation are 
  the emission coefficient, the absorption coefficient, the relative energy shift of the particles with respect to the medium, 
  and the aberration factor, which are all evaluated along the particle geodesics.    
The particle geodesic is obtained by 
  solving the equation of motion of particles for the specified space-time metric 
  with the location of the observer assigned. 
The local values of the emission coefficient and the absorption coefficient can be calculated 
  when the thermodynamic conditions of the medium are given,
  and the relative energy shift and the aberration factor can be determined when the hydrodynamic properties of the medium are known.
   
Here we demonstrate the application of the covariant radiative transfer formulation  
  and calculate the emission from 3D objects in a gravitational field.   
We consider accretion tori around rotating black holes.  
The tori may have several emission components with different optical depths.  
Accretion tori are 3D objects with internal structure. 
They are in contrast to optically thick, geometrically thin accretion disks,   
  which are 2D objects 
  where explicit covariant radiative transfer is unnecessary 
  in determining how the radiation propagates and is modified inside the disk.    

\subsection{Modelling accretion tori} 

For accreting objects, 
the accretion luminosity $L_{acc}$ roughly scales with the mass accretion rate $\dot M$ as $L_{acc} = \varepsilon {\dot M}$.  
The conversion parameter $\varepsilon \sim 10^{20}~{\rm erg \ g}^{-1}$ for black holes.  
The formation of geometrically thin accretion disks around a black hole requires that 
  the radiation pressure force in the accretion flow 
  is much smaller than the local gravitational force exerted by the black hole. 
This condition is usually satisfied 
  when $\dot M$ is sufficiently low,  
  such that $L_{acc}$ is much lower than the Eddington luminosity, 
  which is given by   
  $L_{Edd} = 1.4 \times 10^{38}  ({M}/{M_{\odot}}) \ {\rm erg~s}^{-1}$,  
  where $M$ is the mass of the accretor, 
  and $M_{\odot}$ the solar mass.    
Because $L_{acc}$ increases with $\dot M$, high $\dot M$ implies high $L_{acc}$ and hence a large radiative pressure within the disk  
  where the radiation is liberated.  
When $L_{acc}$ approaches $L_{Edd}$, 
  the radiation pressure force become comparable to the local gravitational force, 
  and the accretion disk thereby inflates and become a torus \citep[see e.g.][]{Frank2002}.   

In general, full knowledge of the fluid viscosity is required 
  in determining the structure and hydrodynamics of the accretion torus.   
However, in accretion tori the angular momentum transport is non-local.   
The process cannot be parametrised with a local viscosity,   
  as in the case of modelling geometrically thin accretion disks,
  where the $\alpha$-viscosity prescription is often used \citep{Shakura1973, Abramowicz1988}. 
It is believed that the angular momentum transport in accretion disks/tori 
  is mediated by tangled magnetic fields that permeate the flow 
  \citep[e.g.\ magneto-rotational instability (MRI), see][]{Hawley2000,Balbus2003}.  
In principle this magnetic viscosity and the flow hydrodynamics need to be determined simultaneously and self-consistently.     
Nevertheless, certain phenomenological prescriptions are proposed to bypass the viscosity calculations, 
  e.g.\  assuming an angular momentum distribution within the torus instead of solving for the distribution.  
With this, the structure of the accretion torus can be determined by solving only the remaining hydrodynamic equations and the equation of state.  
For the purposes of this work we considered this phenomenological approach   
  and constructed accretion tori assuming a specific angular velocity profile within the torus 
  \citep[see also][]{Fuerst2004,Fuerst2007} and \cite{Abramowicz2005}.  
With the angular velocity profile specified, we determined the density and entire flow profiles in terms of certain normalised variables.   
The resulting accretion torus was then rescaled,  
  using the results from accretion tori/disks obtained by numerical MRI simulations (\cite{Hawley2000}; \cite{Balbus2003}).  
The torus model constructed as such is able to capture the geometrical aspects of the MRI accretion tori 
  and the physical conditions within the accretion flow.

\subsection{Emission surface of rotationally supported torus}
\label{sub}

To  calculate the emission from an opaque accretion torus, 
  we need only to specify the torus' boundary surface and its physical conditions.  
The simplest model that allows us to specify  a boundary emission surface is a rotationally supported torus. 
In it the total pressure force is balanced by the 4-acceleration in an arbitrary fluid element.     
The inner boundary of the torus is defined where this balance breaks down. 
With an appropriate parameterisation of the angular velocity profile  ($\Omega$, as a function of position in the torus) 
  we can derive the 4-acceleration and hence obtain the pressure force.      
Tracing the isobars gives the isobaric surfaces in the torus.  
The process is essential in constructing the gradient contours of the 4-acceleration in the $r-\theta$ plane (in the $(t, r,\theta,\phi)$ spherical co-ordinate system).  
The torus boundary surface is simply the outermost allowed isobaric surface.


\begin{figure}[htbp]
\begin{center} 
  \includegraphics[width=0.47\textwidth]{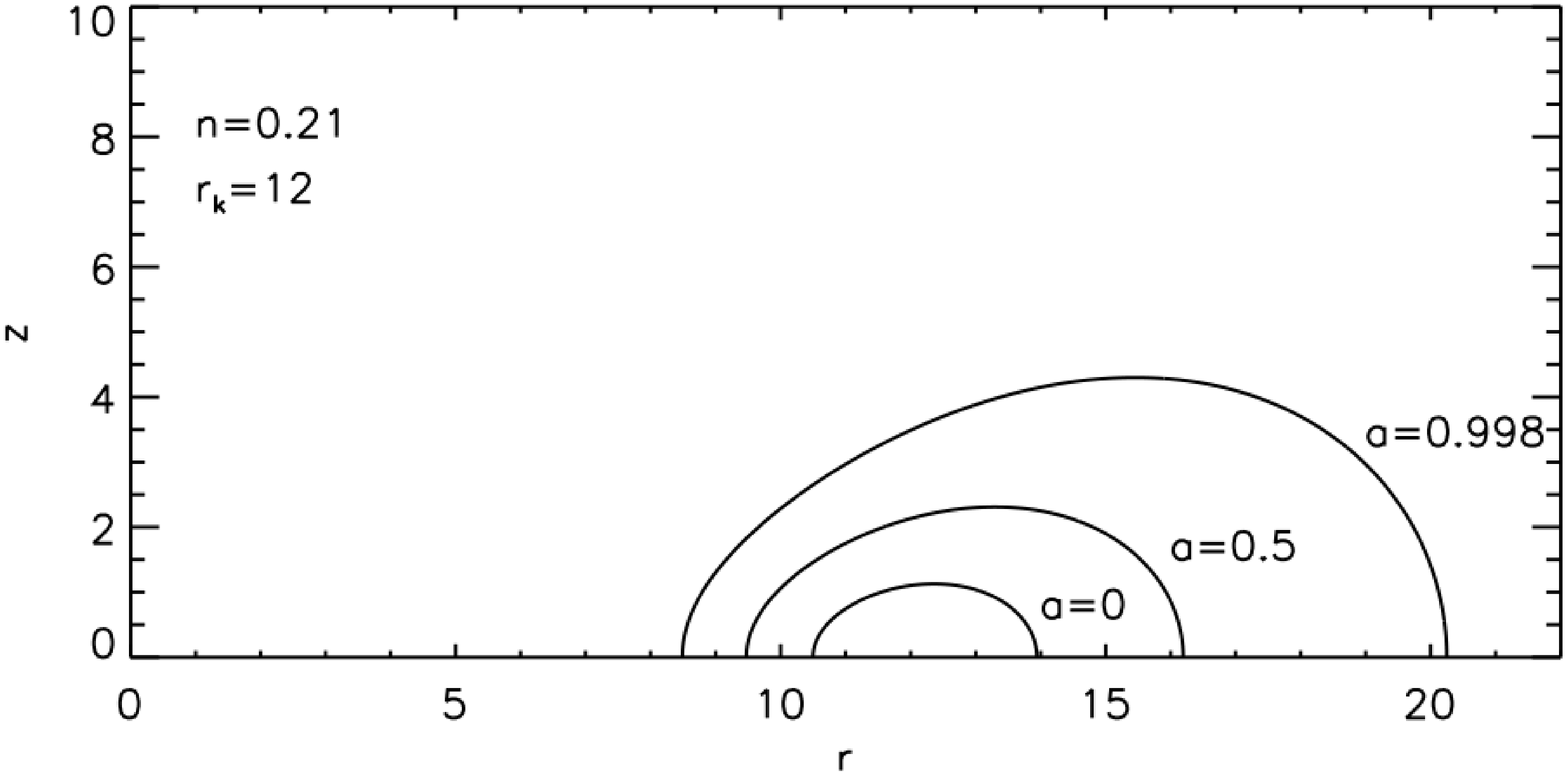}  \\  
  \includegraphics[width=0.47\textwidth]{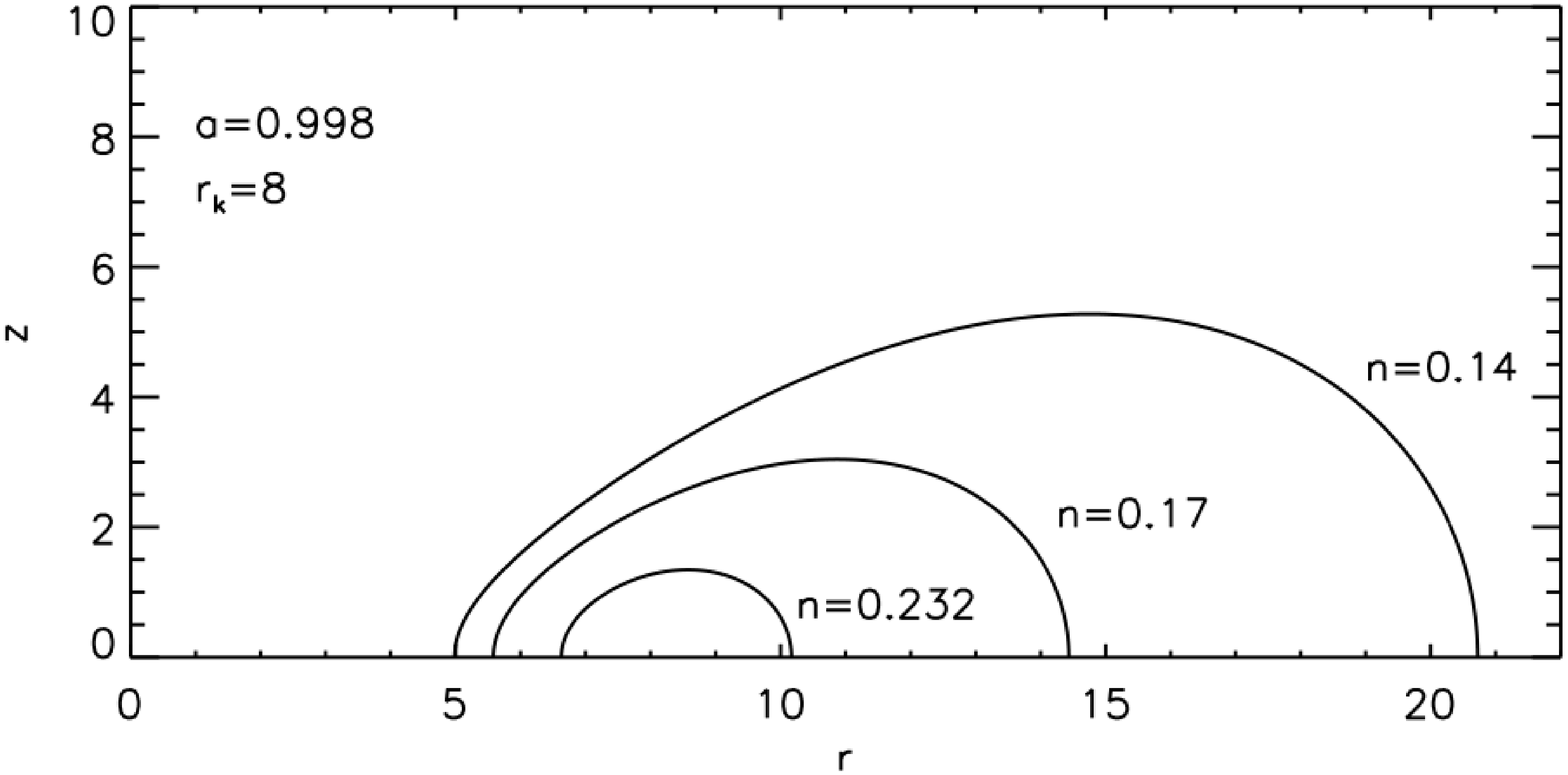} 
\caption{Cross-sections of the boundary surfaces of model rotationally supported tori
  in cylindrical coordinates with $z =  r\cos\theta$ (the equatorial plane of the tori is $z = 0$).   
The top panel shows the tori around Kerr black holes with spin parameters $a= 0$, 0.5 and 0.988. 
The angular velocity profile index of the tori $n= 0.21$, and the Kelperian radius  $r_{k} = 12~r_{\rm g}$. 
The bottom panel shows the tori with angular velocity indices $n = 0.14$, 0.17 and 0.232.  
The Keplerian radius of the tori is $r_{k} = 8$, 
  and the black-hole spin parameter $a = 0.998$. } 
\end{center}
\label{fig-1}
\end{figure} 



We consider a stationary, axisymmetric, rotationally supported accretion torus.   
The symmetry axis of the torus is aligned with the spin vector of the central black hole.  
The 4-acceleration of the flow is derived from the 4-velocity 
\begin{equation} 
a^{\alpha} = u^{\alpha},_{\beta}u^{\beta}+\Gamma^{\alpha}_{\beta\sigma}u^{\beta}u^{\sigma} \ , 
\end{equation}  
  where $\Gamma^{\alpha}_{\beta\sigma}$ is the Christoffel symbol. 
Because the poloidal components of the flow are dynamically unimportant, 
  we set the 4-velocity as $u^{\alpha}=(u^{t},0,0,u^{\phi})$.  
It follows that $u^{\alpha},_{\beta}u^{\beta} = 0$ and  
\begin{equation}  
a^{\alpha} = \Gamma^{\alpha}_{tt}u^{t}u^{t}+2 \Gamma^{\alpha}_{t\phi}u^{t}u^{\phi} +\Gamma^{\alpha}_{\phi\phi}u^{\phi}u^{\phi} \ .  
\end{equation}  
In Boyer-Lindquist co-ordinates, 
   the ${r}$ and ${\theta}$ components of the 4-acceleration are 
\begin{eqnarray}
    a_{r} & = & \frac{\Sigma}{\Delta}a^{r} = -\left[ M\left(\frac{\Sigma-2r^{2}}{\Sigma^{2}}\right)\Big(\dot{t}-a\sin^{2}\theta \ \dot{\phi} \Big)^{2}+r\sin^{2}\theta \ \dot{\phi}^{2} \right]\  , \\
    a_{\theta} & = & \Sigma \ \! a^{\theta}= -\sin 2\theta \  \bigg(\frac{Mr}{\Sigma^{2}}\big[a\dot{t}-\big(r^{2}+a^{2}\big)\dot{\phi}\big]^{2}+\frac{\Delta}{2} \dot{\phi}^{2} \bigg)\  , 
\end{eqnarray} 
   where $\Delta = r^{2}-2Mr+a^{2}$, $\Sigma = r^{2}+a^{2}\cos^{2}\theta$, 
   and $M$ and $a$ are respectively the mass and spin parameters of the central gravitating body. 
Setting $a_{\alpha}u^{\alpha}=0$ yields a set of differential equations for the isobaric surfaces. 
It is more convenient to express the equations as 
\begin{eqnarray}
    \frac{dr}{d\xi} &=& \frac{\psi_{2}}{\sqrt{\psi_{2}^{2}+\Delta\psi_{1}^{2}}}\ , 
    \label{surface-1} \\ 
    \frac{d\theta}{d\xi} &=& \frac{-\psi_{1}}{\sqrt{\psi_{2}^{2}+\Delta\psi_{1}^{2}}}\ ,
\label{surface-2}  
\end{eqnarray} 
 for the calculation and construction of the torus boundary surface. 
Here $\xi$ is an auxiliary variable, and the functions $\psi_{1}$ and $\psi_{2}$ are given by  
\begin{eqnarray}
  \psi_{1} &=& M\bigg(\frac{\Sigma-2r^{2}}{\Sigma^{2}} \bigg)\Big(\Omega^{-1}-a\sin\theta \Big)^{2}+r\sin^{2}\theta \ , \\
  \psi_{2} &=& \sin 2\theta  \  \bigg(\frac{Mr}{\Sigma^{2}}\big[a \ \! \Omega^{-1}-\big(r^{2}+a^{2}\big)\big]^{2}+\frac{\Delta}{2}\bigg)\ .
\end{eqnarray}  
Solving equations (\ref{surface-1}) and (\ref{surface-2}), 
  with the inner boundary radius of the torus and the specified angular velocity profile 
  gives the torus boundary surface.   
  
We consider an angular velocity profile with the form 
\begin{equation}
    \Omega(r\sin \theta) = \frac{\sqrt{M}}{(r \ \! \sin\theta)^{3/2}+a\sqrt{M}}\bigg(\frac{r_{k}}{r \ \!\sin \theta} \bigg)^{n}, \   
\label{para}
\end{equation}  
  following \cite{Fuerst2004,Fuerst2007}, 
  where $r_{k}$ is the radius on the equatorial plane at which the material circulates with a Keplerian velocity.  
The differential rotational velocity gradient gives rise to an implicit pressure force, 
  supporting the torus material above and below the equatorial plane.  
The parametrisation using the variable $r \ \! \sin\theta$ 
  ensures the constant density and pressure surfaces coincide in the Newtonian limit  
  and a polytropic equation of state is applicable for the flow.  
The index parameter $n$ is crucial for regulating the pressure forces, thus adjusting the torus' geometrical aspect ratio.   
Its property is similar to that of the $q$ index of the von Zeipel parameter in the study of stability of accretion disks \citep[see][]{Chakrabarti1985,Blaes1988}.   
Generally, $n\approx q-1.5$, with the relation being exact for Schwarzschild black holes. 
Tori with $q >\sqrt{3}$ are unstable. In the Newtonian limit tori with $n=0.232$ are marginally stable.

The angular velocity and angular momentum of the flow are   
  $\Omega=u^{\phi}/u^{t}$ and $l=-u_{\phi}/u_{t}$ respectively, 
  and the redshift factor is given by     
\begin{eqnarray}
  A  & =  &  u^{t}   
      =   \left [-( g_{tt}+2\Omega g_{t\phi}+\Omega^{2}g_{\phi \phi})\right]^{-1/2}\   ,  
\label{a-eqn}
\end{eqnarray}    
  and the energy per unit inertial mass of the flow material is
\begin{eqnarray}
  U &  = &  -u_{t}  
        =  - \sqrt{\frac{g_{t\phi}^{2}-g_{tt}g_{\phi\phi}}{g_{\phi\phi}+2 l g_{t\phi}+l^{2}g_{tt}}}  \ .  
\label{u-eqn}    
\end{eqnarray}       
These two quantities are related via 
\begin{equation} 
AU = \frac{1}{(1-l \Omega)} \ . 
\label{au} 
\end{equation}
Zero values for the denominators in equations (\ref{a-eqn}) and (\ref{u-eqn}) correspond to  
  the locations (the photon surface) where the local flow speeds reach the speed of light.    
  
In this model we need to specify the condition for the inner boundary radius. 
We take it as the intersection of the isobaric surface 
  with either the orbits of marginal stability or the limiting surface of photon orbits, whichever has a larger radius.     
Usually the photon surface is within the marginally stable orbit.  
The inner boundary of the torus is therefore in general determined by the outermost surface of the torus, 
  which satisfies ${\partial U}/{\partial r}=0$. 
Solving this gives 
\begin{eqnarray}
    2aM \sin^{4}\theta\bigg[\frac{r^{2}}{\Sigma}-\Big(r^{2}+a^{2}+\frac{a^{2}Mr\sin^{2}\theta}{\Sigma} \Big)\frac{\Sigma-2r^{2}}{\Sigma^{2}} \bigg]\Omega^{3}  
       & & \nonumber \\ 
   & &\hspace*{-7.6cm}  +  \sin^{2}\theta\bigg[M\bigg(\frac{6Mr(r^{2}+a^{2})}{\Sigma}+3\Delta-\Sigma \bigg)
        \frac{\Sigma-2r^{2}}{\Sigma^{2}}+r\bigg(1-\frac{2Mr}{\Sigma} \bigg) \bigg]\Omega^{2} \nonumber   \\
   & &\hspace*{-7.6cm} -\frac{6aM^{2}r\sin^{2}\theta}{\Sigma}\bigg(\frac{\Sigma-2r^{2}}{\Sigma^{2}} \bigg)\Omega \nonumber  \\
   & &\hspace*{-7.6cm} +\Delta\sin^{2}\theta \ \! \Omega \frac{\partial \Omega}{\partial r}-M\bigg(1-\frac{2Mr}{\Sigma} \bigg)\frac{\Sigma-2r^{2}}{\Sigma^{2}}=0 \ ,  
\end{eqnarray}  
   which when solved for $r$, yields the inner edge of the torus (for a particular choice of $r_{k}$ and $n$). 
Equations (\ref{surface-1}) and (\ref{surface-2}) for the torus surface are now readily integrated.       

Figure~1 shows the boundary surfaces of  rotationally supported tori with various system parameters.    
The shape of the tori is determined by the rotational velocity index of the torus $n$ and the black-hole spin parameter $a$.  
When $r_{k}$ is fixed, 
  the vertical thickness of the torus increases with $a$ but decreases with $n$. 
The degree of self-occulation of an optically thick torus   
   and hence the spectral properties of the emission  
   depend on the aspect ratio of the torus and the viewing inclination.   
Note that the tori in Figure~1 are purely rotation-supported. 
The thermal pressure of the torus gas and the radiative pressure of the emission from the gas 
  have not been included in their construction.      
The presence of gas pressure and radiation pressure will modify the aspect ratio of the tori. 
In the next subsection we will consider a more general situation,  
  which includes the gas pressure and radiative pressure,   
  and construct tori with internal density and temperature structures.

\subsection{Pressure supported torus structure}  

 

Accretion tori resemble stars that have an atmosphere with an optical depth gradient. 
While emission from an opaque torus stems from an unobscured skin (surface) layer of the torus,  
  emission from a translucent or an optically thin accretion torus is formed by the emission contribution from all regions within the torus.  
The thermodynamic and hydrodynamic structures of the torus determine the spectral properties of its emission,  
  and so they must be determined prior to performing the radiative transfer calculations.  

To model the internal structure of the accretion tori,  
  we adopted a prescription given by \cite{Abramowicz1978} and \cite{Kozlowski1978}.  
We considered the tori as stationary and axisymmetric.  
They consist of a perfect fluid, 
  and the stress-energy-momentum tensor of the flow is given by 
\begin{equation} 
  T^{\alpha \beta}=(\rho+{P}+\epsilon)u^{\alpha}u^{\beta}+ {P}g^{\alpha \beta} \ ,
\end{equation}
  where $P$ is the pressure, $\rho$ is the density, and $\epsilon$ is the fluid internal energy. 
Because $T^{\alpha \beta}_{\ \ \ ;\beta}=0$,  we have 
\begin{equation}
   (\rho+P+\epsilon)_{,\beta}u^{\alpha}u^{\beta}+(\rho+P+\epsilon)(u^{\alpha}_{\ ;\beta}u^{\beta}+u^{\alpha}u^{\beta}_{\ ;\beta})+{P}_{,\beta}g^{\alpha \beta}=0 \ .
\end{equation}
Projecting perpendicular to the velocity with the projection tensor $P^{\alpha \beta}$ yields the momentum equation 
\begin{equation}
   (\rho+{P}+\epsilon)u^{\alpha}_{\ ;\beta}u^{\beta}+{P}_{,\beta}g^{\alpha \beta}=0\ . 
\label{mom-eq}
\end{equation} 

Because the torus is stationary and axisymmetric, it has negligible poloidal flow components. 
Hence,    
\begin{equation}
   u_{\alpha ;\beta}u^{\beta} =  -\Gamma^{\sigma}_{\alpha \beta}u_{\sigma}u^{\beta} 
    = -\frac{1}{2}u^{\sigma} u^{\beta}g_{\sigma \beta, \alpha} \ .
\end{equation}
Differentiating $u^{\alpha}u_{\alpha}=-1$ gives 
\begin{equation}
   (u^{\alpha} u_{\alpha})_{,\delta} 
                         = g_{\alpha \beta , \delta}u^{\alpha}u^{\beta}+2u^{\alpha}u_{\alpha,\delta} \nonumber  = 0 \ . 
\end{equation}
Thus, 
\begin{equation}
   u^{\alpha} u_{\alpha,\delta} = -\frac{1}{2}u^{\alpha}u^{\beta}g_{\alpha \beta ,\delta}\ .
\end{equation}
It follows that
\begin{equation}
  u_{\alpha ; \beta}u^{\beta}=u^{\beta}u_{\beta , \alpha}  = u^{t}\partial_{\alpha}u_{t}+u^{\phi}\partial_{\alpha}u_{\phi}\   , 
\label{uu}
\end{equation}
   where $\partial_{\alpha}$ is the gradient in the $x^{\alpha}$ direction.  
Recall that  $\Omega=u^{\phi}/u^{t}$ and $l=-u_{\phi}/u_{t}$.  
Therefore, we have 
\begin{equation}
   u^{\phi}u_{t}=  \Omega ( u^{t}u_{t}) =  -\frac{\Omega}{1-l\Omega}\  .
\end{equation}
The gradient of $l$ is  
\begin{equation}
   \partial_{\alpha}l=\frac{u_{\phi}}{u_{t}^{2}}\partial_{\alpha}u_{t}-\frac{1}{u_{t}}\partial_{\alpha}u_{\phi}\  .
\end{equation}
It follows that 
\begin{equation}
   \frac{\Omega\partial_{\alpha}l}{1-l\Omega}=\frac{1}{u_{t}}\partial_{\alpha}u_{t}+u^{t}\partial_{\alpha}u_{t}+u^{\phi}\partial_{\alpha}u_{\phi}\ .
\end{equation}
Equation (\ref{uu}) can now be expressed as 
\begin{equation}
   u_{\alpha ; \beta}u^{\beta}=\frac{\Omega\partial_{\alpha}l}{1-l\Omega}-\frac{1}{u_{t}}\partial_{\alpha}u_{t}\  . 
\end{equation}
Thus, 
\begin{equation}
   \frac{\partial_{\alpha}P}{\rho+P+\epsilon}=\partial_{\alpha}\ln (u_{t})-\frac{\Omega\partial_{\alpha}l}{1-l\Omega} 
\end{equation}
   (cf. equation (7) in \cite{Abramowicz1978} for the accretion torus).
   
An equation of state is required to close the system of equations and the gas within the torus is assumed to be barotropic. 
Owing to the complexity of the Kerr metric, the solution must be computed numerically. 
However, in the special case of $l=\textrm{constant}$, corresponding to a marginally stable torus, 
  there is an analytic solution 
\begin{equation}
   \int_{0}^P \frac{dP'}{\rho+P'+\epsilon}=\ln (u_{t})-\ln (u_{t})_{inner}\  ,
\end{equation}
   where $\ln (u_{t})_{inner}$ is evaluated at the inner edge of the torus.

  
\begin{figure}[htbp]
\begin{center} 
     \vspace*{0.2cm} 
  \includegraphics[width=0.4\textwidth]{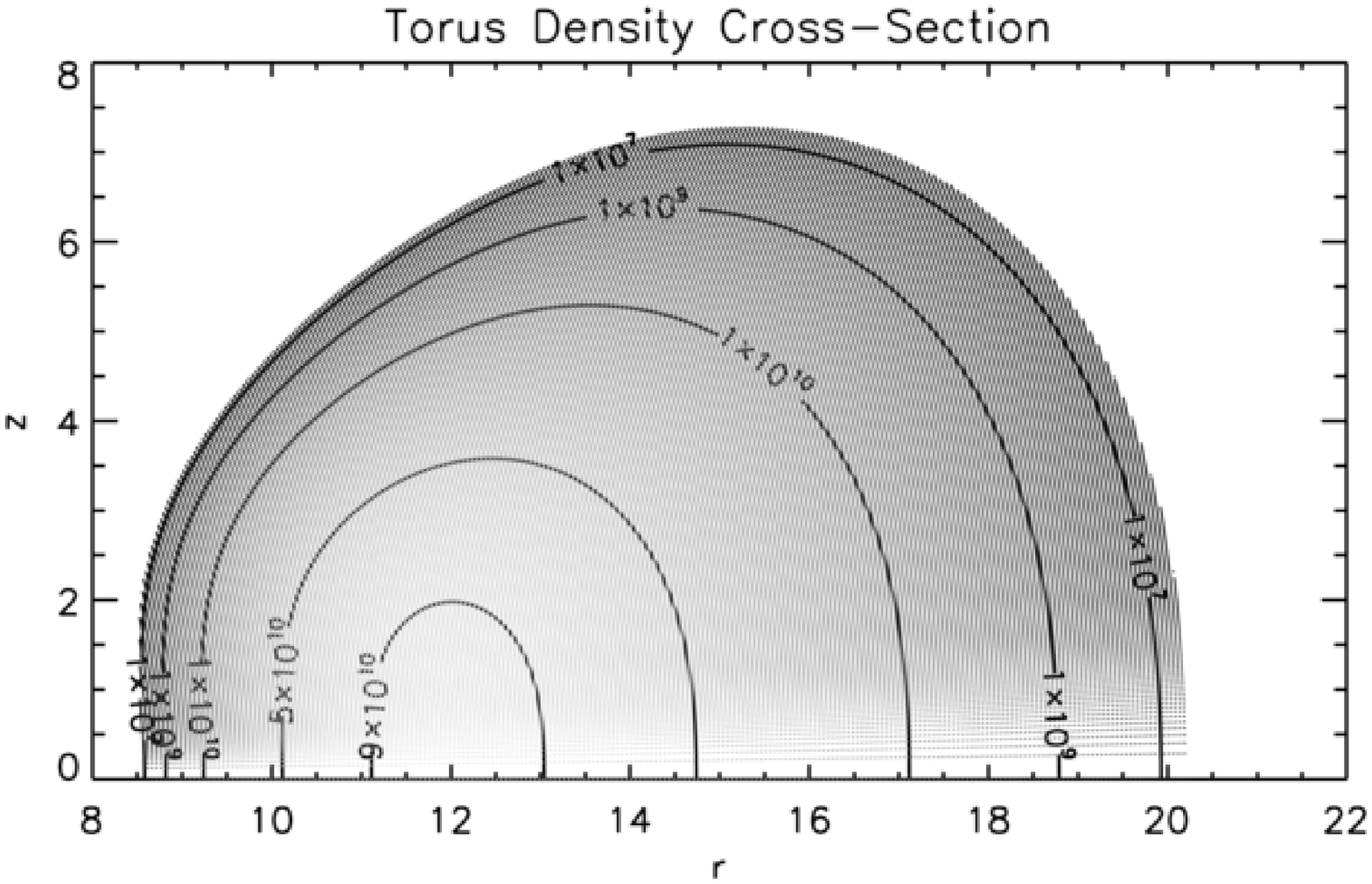}  \\  
  \vspace*{0.2cm} 
  \includegraphics[width=0.4\textwidth]{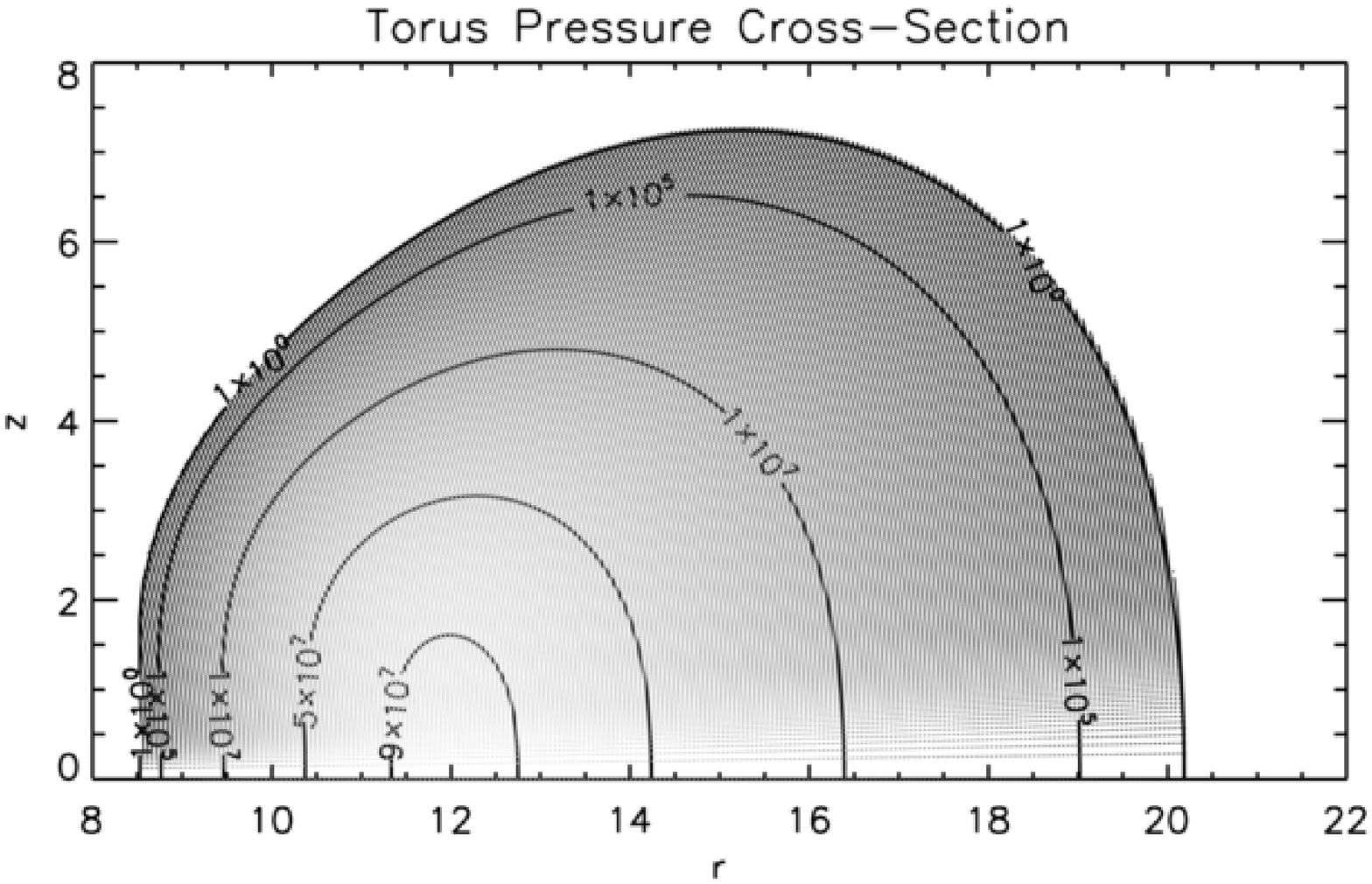}  \\  
    \vspace*{0.2cm} 
  \includegraphics[width=0.4\textwidth]{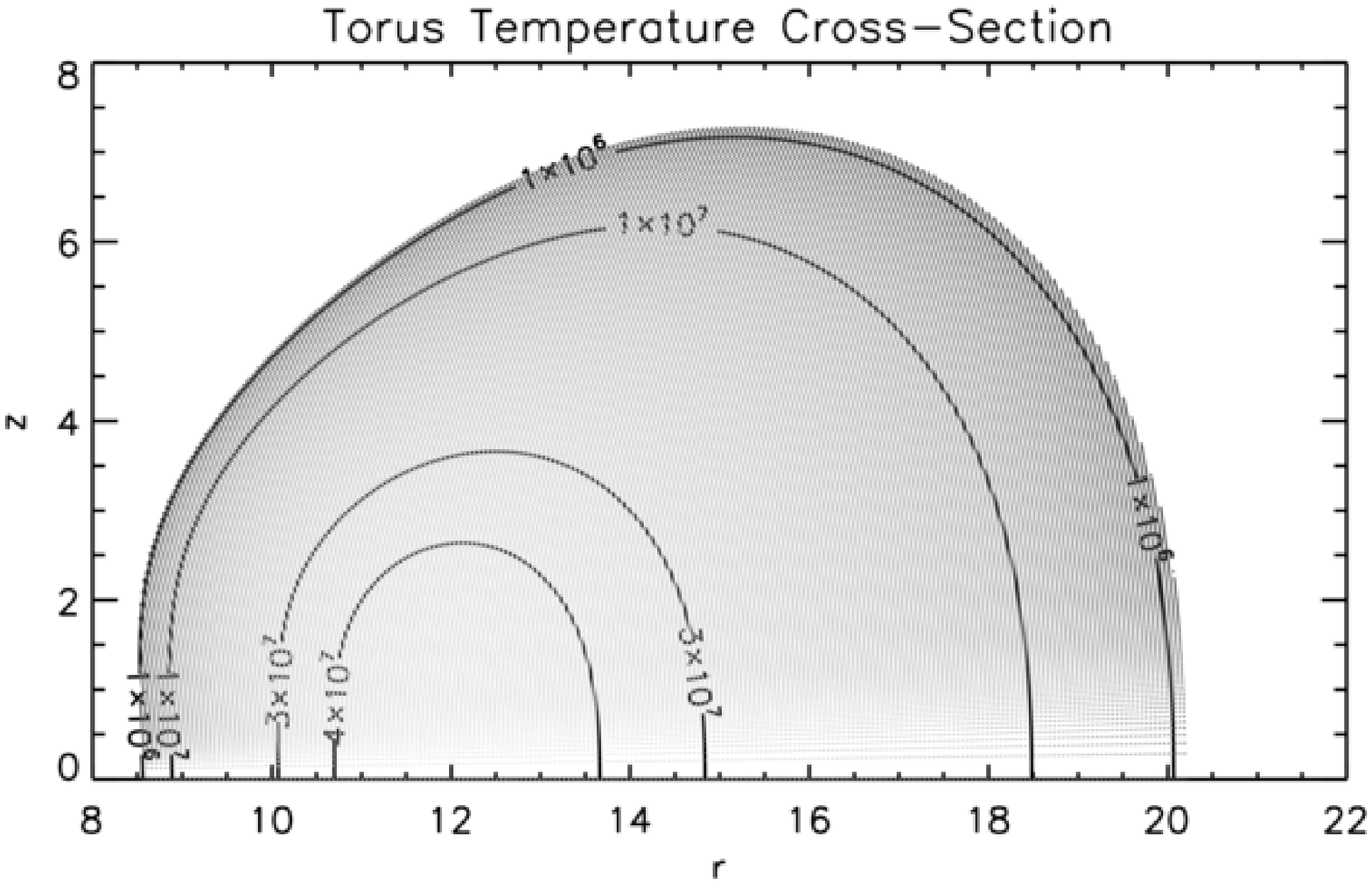}  
\caption{Cross-sections show the density, pressure and temperature contours  
  in model translucent tori (panels from top to bottom). 
 The angular velocity profile index of the tori $n = 0.21$, 
   the Keplarian radius $r_{k} = 12~r_{\rm g}$. 
 The black-hole spin paramter $a = 0.998$.  
 The central density of the torus $\rho_{c} = 10^{11}~{\rm cm}^{-3}$. 
 The ratio of the gas pressure to the total pressure $\beta = 1.235 \times 10^{-5}$.} 
\end{center}
\label{fig-2}
\end{figure} 


The transparency of the torus to radiation requires that 
    the local emissivity and opacity within the torus must be specified explicitly. 
As such, the velocity, temperature and density structure of the torus must be determined 
    prior to the radiative transfer calculations.
   
The total pressure within the torus is the sum of the gas pressure and the radiation pressure, i.e. \ 
  $P= P_{gas}+ P_{rad}$,  where
\begin{eqnarray}
   {P}_{gas}&  = &  \frac{\rho k T}{\mu m_{H}} = \beta {P} \ , \\
   {P}_{rad} & = &  \frac{4\sigma}{3c}T^{4} = (1-\beta) {P}\ . 
\end{eqnarray} 
Here $k$ is the Boltzmann constant, $\mu$ the mean molecular weight, $m_{H}$ the mass of hydrogen, 
 $\beta$ the ratio of gas pressure to total pressure, and $\sigma = {\pi^{2}k^{4}}/{60\hbar^{3}c^{2}}$ is the black-body emittance constant.
Eliminating $kT$ in the above equations yields
\begin{equation}
   {P}=\hbar c \bigg[\frac{45(1-\beta)}{\pi^{2}(\mu m_{H}\beta)^{4}} \bigg]^{1/3}\rho^{4/3} \ . 
\label{poly}   
\end{equation} 
For a polytropic equation of state ${P} = \kappa \rho^{\Gamma}$,  
  the internal energy is related to the pressure by $\epsilon = {P}/{\Gamma -1}$.
Equation (\ref{poly}) implies $\Gamma=4/3$, and 
$\kappa = \hbar c[{45(1-\beta)}/{\pi^{2}(\mu m_{H}\beta)^{4}}]^{1/3}$.

Combining the momentum equation (\ref{mom-eq}) for a perfect fluid with the polytropic equation of state gives 
\begin{equation}
   \bigg( \rho+\frac{\Gamma}{\Gamma -1}P \bigg)a^{\alpha}=- P_{,\beta}g^{\alpha \beta} \ .
\end{equation}
Differentiating the polytropic equation of state yields
\begin{equation}
   \partial_{\alpha} {P} =\kappa \Gamma \rho^{\Gamma-1} (\partial_{\alpha} \rho)\  .
\end{equation} 
The density structure of the torus is then given by 
\begin{equation}
   \partial_{\alpha} \rho=-a_{\alpha}\bigg(\frac{\rho^{2-\Gamma}}{\kappa \Gamma}+\frac{\rho}{\Gamma-1} \bigg). 
\label{rho}
\end{equation}
Introducing a new variable $\xi$ (akin to logarithm of temperature), where
$\xi = \ln (\Gamma -1+\Gamma \kappa \rho^{\Gamma -1})$,  
equation (\ref{rho}) simplifies to
\begin{equation}
   \partial_{\alpha} \xi=-a_{\alpha} \ .
\end{equation} 
The stationary and axisymmetry conditions imply that 
 there are only two non-trivial components ($r$ and $\theta$) in the equation. 
By evaluating the line integral from  $r=r_{K}$, $\rho=\rho_{c}$ at the torus centre to the required ($r, \theta$) location, 
  the density field $\rho(r,\theta)$ is determined. 
     
If the pressure in the torus is dominated by the radiative pressure, 
  we may set $(1 -\beta) \approx 1$. 
Then we have 
\begin{eqnarray}
  P  & = & \hbar c \left[\frac{45}{\pi^{2}(\mu m_{H}\beta)^{4}}\right]^{1/3}   \rho^{4/3} \  ,    \\ 
   kT  & = &  \hbar c \bigg[\frac{45}{\pi^{2}(\mu m_{H}\beta)} \bigg]^{1/3}\rho^{1/3} \ .
\end{eqnarray}      
With $\rho(r,\theta)$ determined,
   the pressure and temperature are readily calculated. 

Figure~2 shows the density, pressure and temperature structures of a model torus 
  with $n = 0.21$, $r_{k} = 12~r_{\rm g}$, $\beta =1.235 \times 10^{-5}$, and $\rho_{c} = 10^{11}~{\rm cm}^{-3}$, 
  where $r_{\rm g} = GM/c^{2}$ is the gravitational radius of a black hole with a mass $M$, and $G$ is the gravitational constant. 
The spin parameter of the black hole $a= 0.998$    
There are several noticeable features in the torus.  
The torus is pressure supported. 
Its boundary is located where the density $\rho$ and hence the pressure $P$ vanish. 
The temperature $T$ also vanishes at the torus boundary surface for the equation of state that we adopted.   
As shown, the temperature $T \sim 10^{7}~$K in most of the torus interior,  
  but it drops rapidly within a very short length of $\sim 1~R_{\rm g}$ and reaches 0~K at the torus boundary.   
Compared with the rotationally supported tori (Figure~1), 
  the rotationally supported tori are larger in vertical extent. 
The vertical inflation of the torus is caused by the addition of radiative and thermal gas pressure forces.

  
\begin{figure}[htbp]
\begin{center}
  \includegraphics[width=0.23\textwidth]{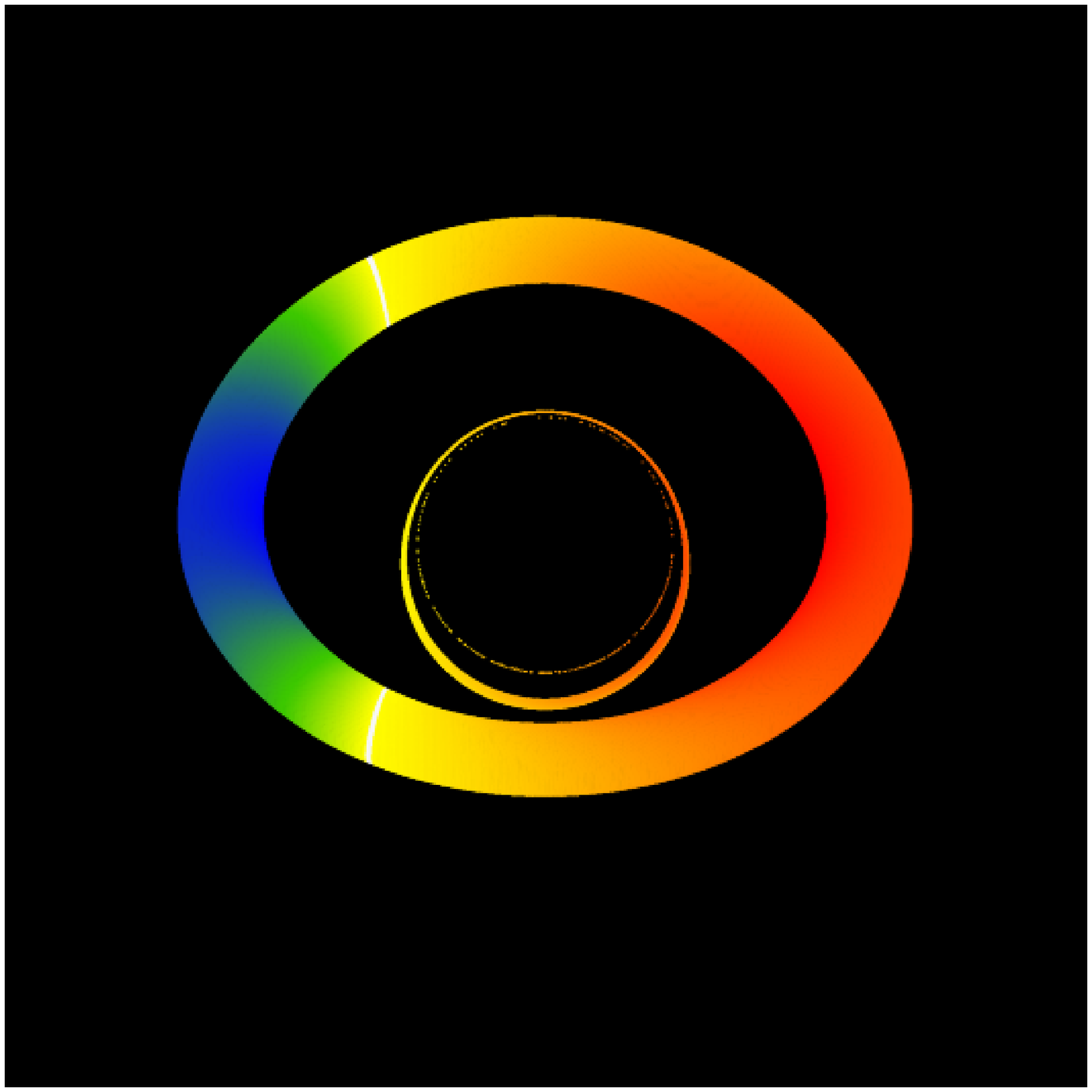}
  \includegraphics[width=0.23\textwidth]{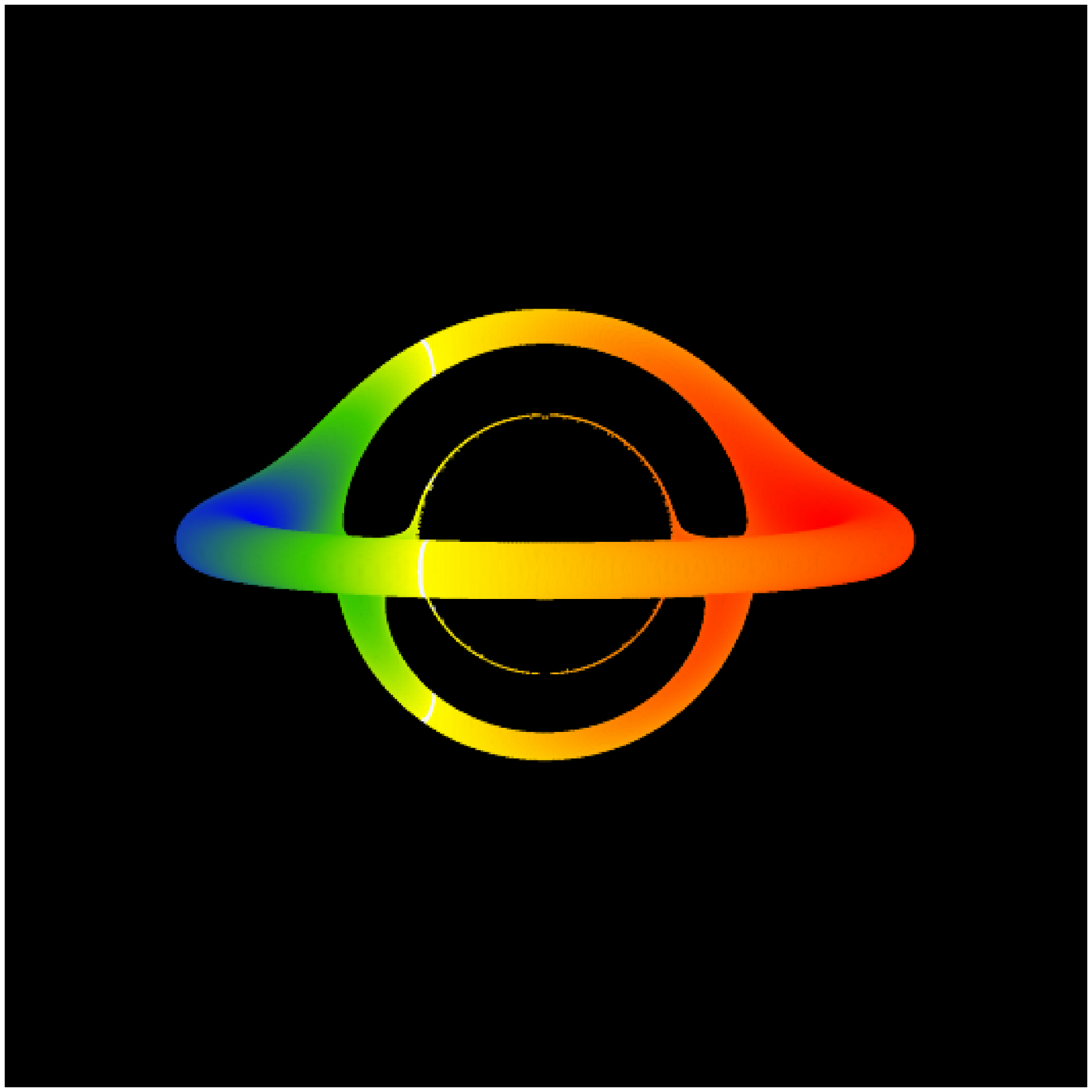}
  \includegraphics[width=0.23\textwidth]{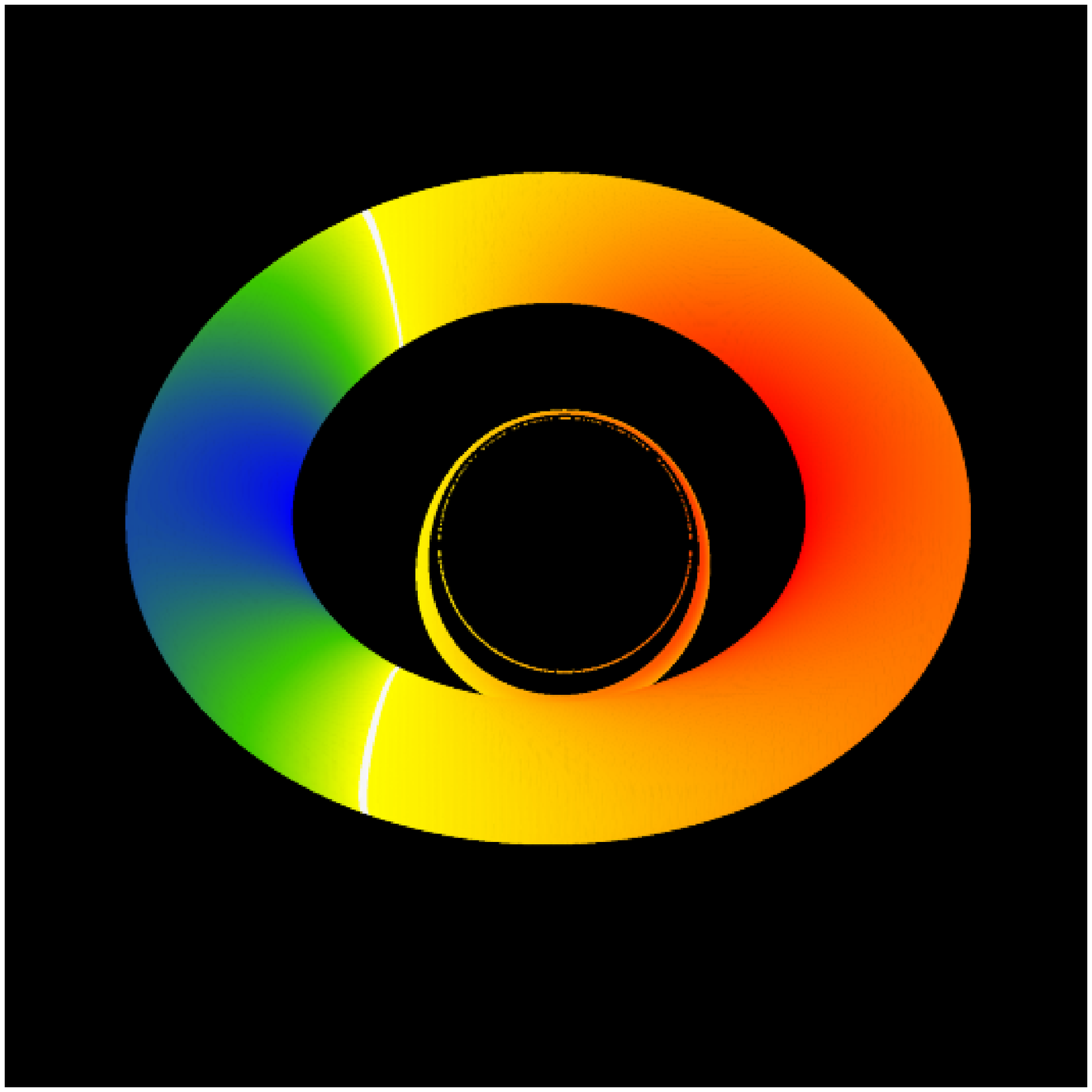}
  \includegraphics[width=0.23\textwidth]{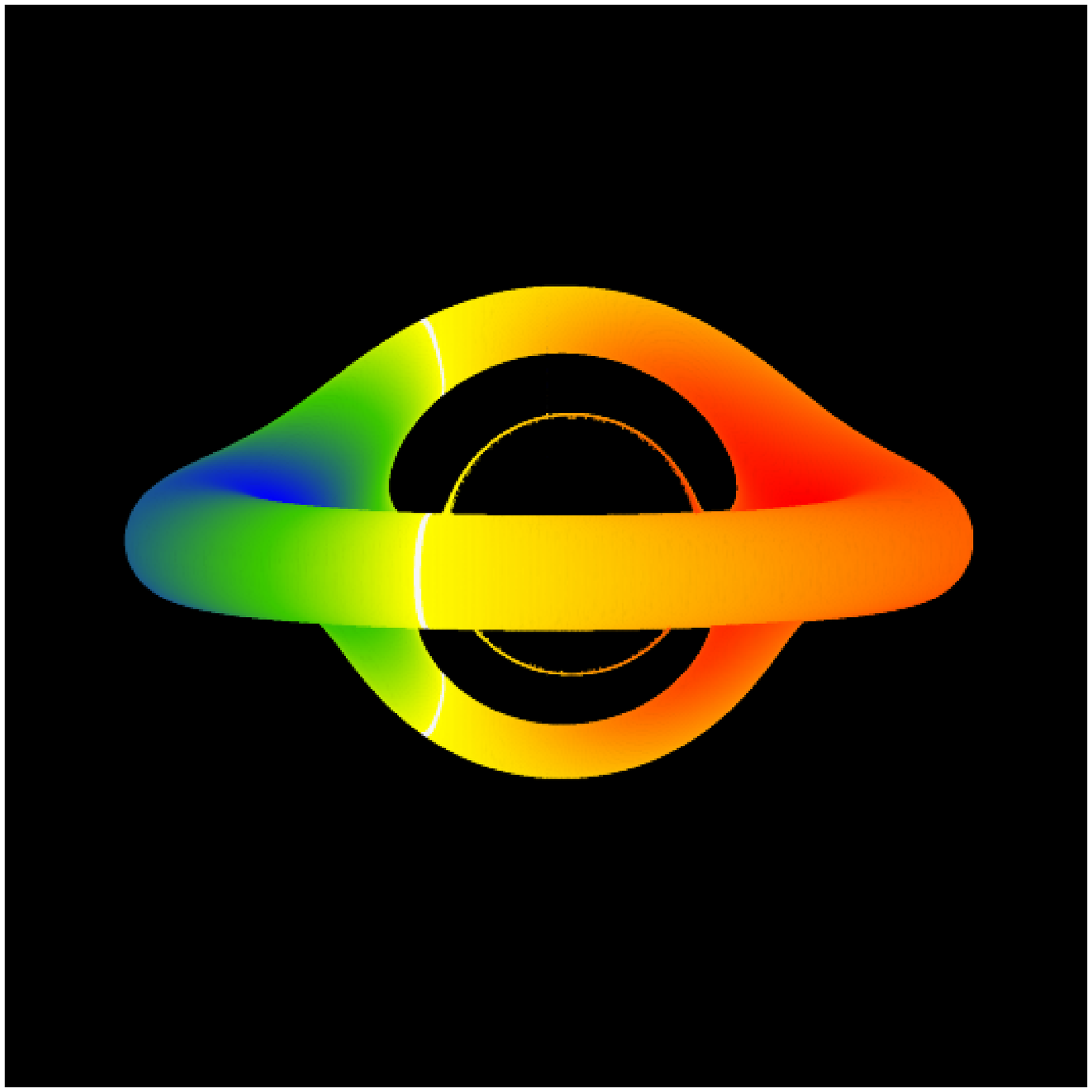}
  \includegraphics[width=0.23\textwidth]{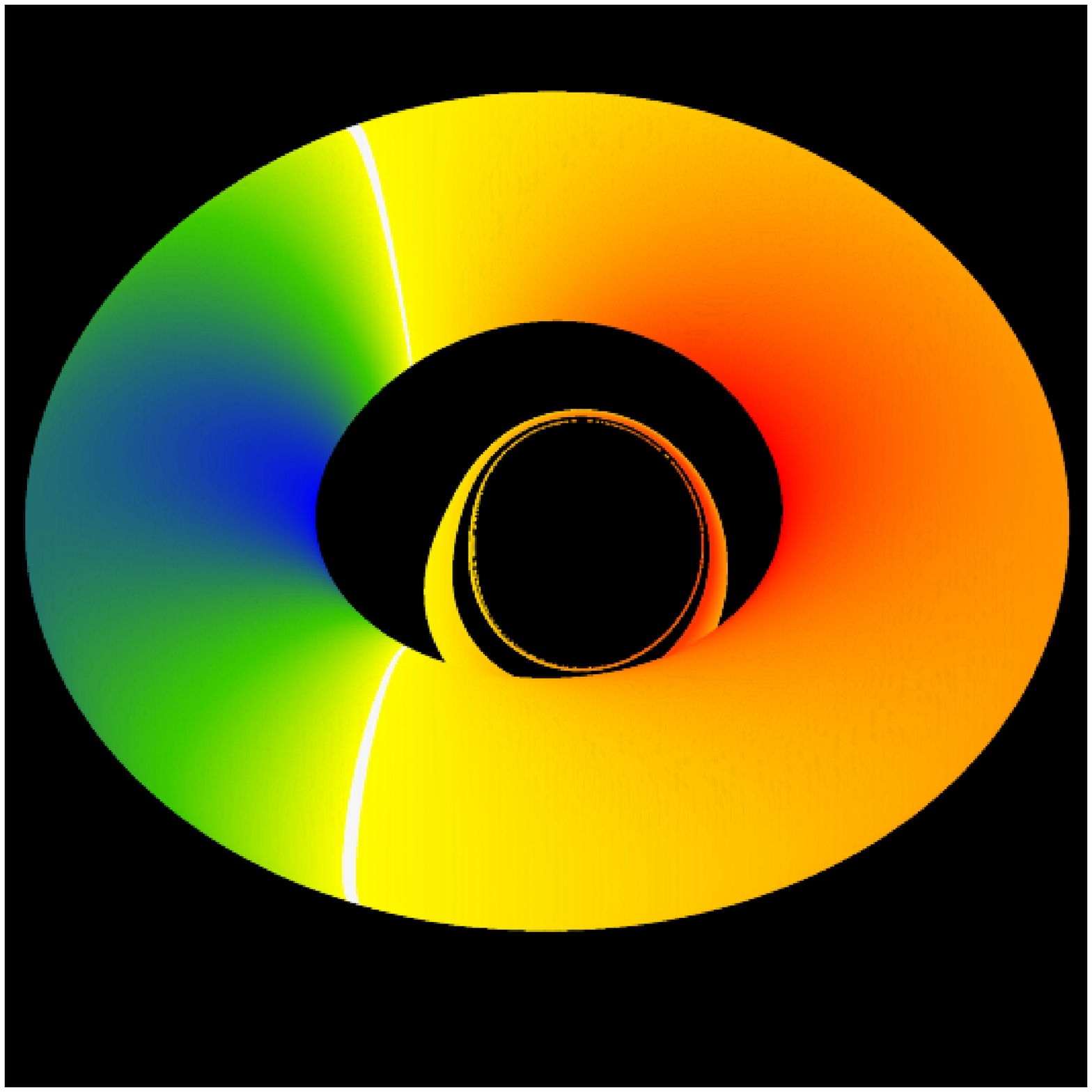}
  \includegraphics[width=0.23\textwidth]{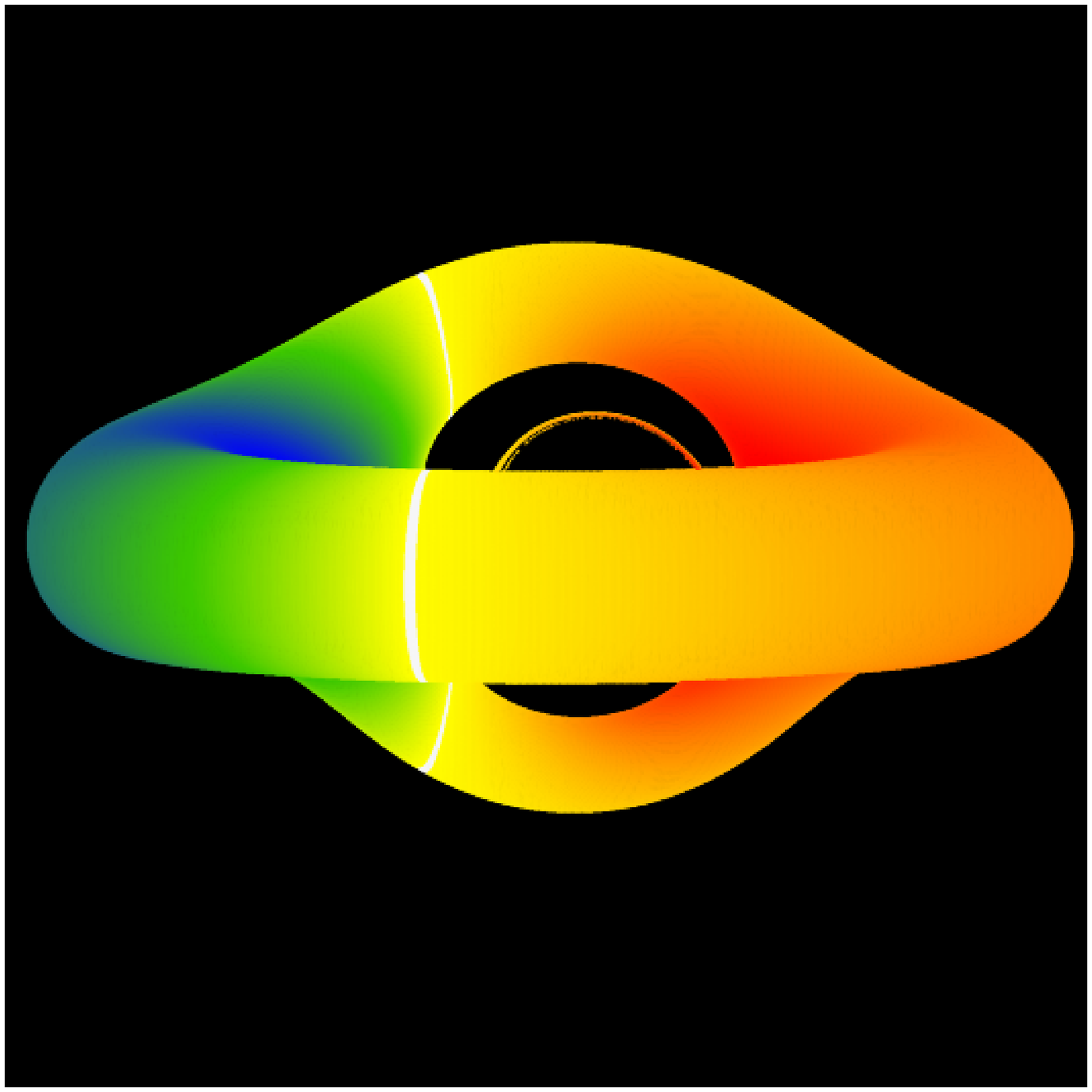}
\caption{False-colour frequency shift maps of the surface emission from opaque tori around different black holes.  
  The  torus parameters are $n=0.232$ and $r_k=12\ r_g$, where $r_{g}$ is the gravitational radius.  
   The black-hole spin parameters are $a = 0$, 0.5 and 0.998 (panels from top to botom). 
   The viewing inclinations of the tori are $45^{\circ}$ (left column) and $85^{\circ}$ (right column). For $a=0$ the range of frequency shifts $\mathrm{E}/\mathrm{E}_{0}$ for $i=45^{\circ}$ and $i=85^{\circ}$ are $(0.874,1.445)$ and $(0.756,1.560)$, respectively . Similarly, for $a=0.5$ the frequency shift ranges are $(0.870,1.487)$ and $(0.755,1.591)$. Finally, for a=0.998 the corresponding frequency shift ranges are $(0.864,1.535)$ and $(0.767,1.616)$.}
\end{center} 
\label{fig-3} 
\end{figure}


\begin{figure}[htbp]
\begin{center} 
   \vspace*{0.2cm} 
   \includegraphics[width=0.425\textwidth]{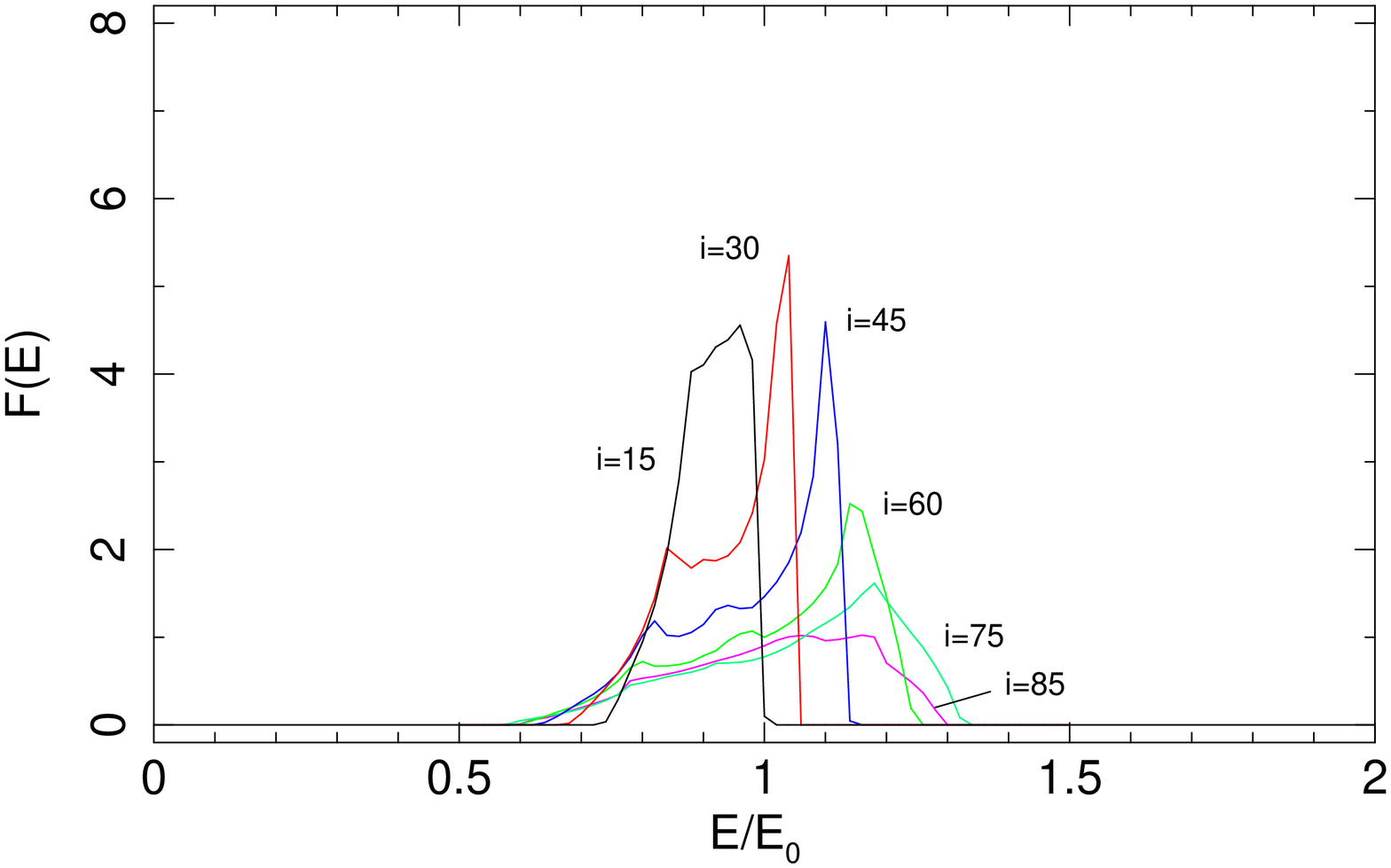}  \\  
   \vspace*{0.12cm} 
   \includegraphics[width=0.425\textwidth]{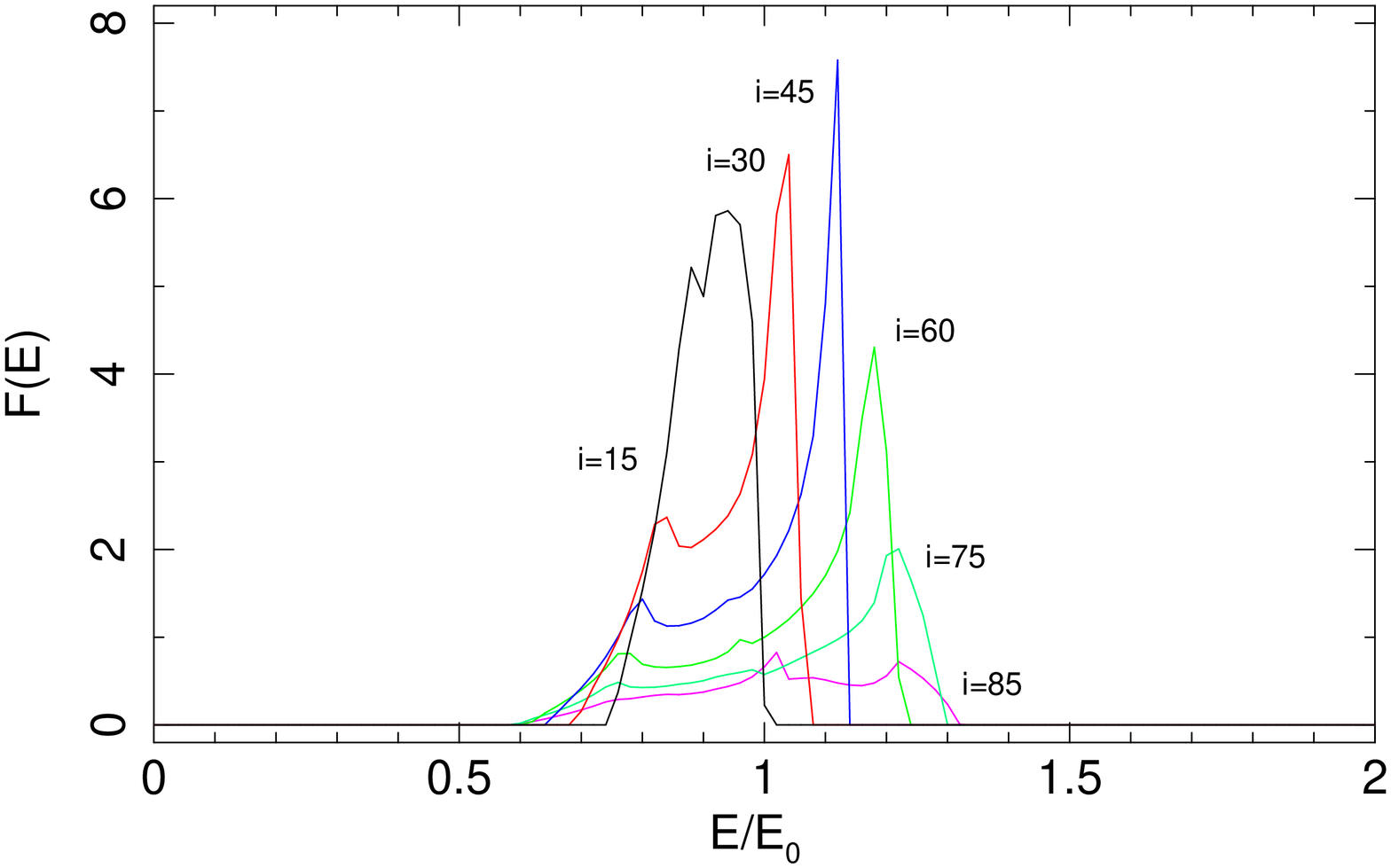}  \\  
\caption{Profiles of emission lines from an opaque rotationally supported accretion torus (top panel)  
   and a geometrically thin optically thick accretion disk (bottom panel) viewed at different inclinations $i$.  
 The torus dynamical parameters are $n = 0.21$ and $r_{k} = 12\ r_{g}$.       
 The inner boundary radius of the torus $r_{in} = 8.486\ r_{g}$ and the outer boundary radius $r_{out} = 20.246\ r_{g}$.   
 The disk has the same values for the inner boundary radius and the outer boundary radius as the torus.  
 In both cases, the black-hole spin parameter $a = 0.998$. 
 The line emissivity is proportional to $r^{-2}$, where $r$ is the radial distance from the central black hole.     
 The line profiles are normalised such that the flux $F(E_{0}) = 1$ 
   at the view inclination angle  $i = 60^{\circ}$,  where $E_{0}$ is the rest-frame line centre energy.    
  } 
\end{center}
\label{fig-4}
\end{figure} 


\section{Radiative transfer calculations for accretion tori}    

\subsection{Emission from opaque rotationally supported tori}     

For the opaque tori, the emission spectrum can be calculated 
  from the emissivity distribution on the torus' boundary surface, 
  with corrections for the relativistic shifts with respect to the distant observer.  
We show in Figure~3 images of rotationally supported tori  
   viewed at inclination angles of $45^{\circ}$ and $90^{\circ}$  
   for various black hole spin parameters. 
The torus is left-right symmetric in shape if the black hole is not rotating.  
The black hole's rotation drags the surrounding space-time around the black hole, 
   and so the torus around a Kerr black hole 
  no longer has a left-right symmetrical shape. 
Optically thick tori would suffer self-eclipsing at high viewing inclination angles. 
Only the unobscured regions on the torus surface  
  would contribute to the emission detectable by the distant observer.   
Each pixel in the torus images shown in Figure~3  
  has only one single value for the relative frequency shift 
 between the emission surface element and the observer.  
We coded this relative frequency shift with colours in the torus images.  
We can easily see that the regions with the highest redshifts and with the highest blueshifts 
   are obscured by the front limb of the torus 
   when the viewing inclination angle is close to $90^{\circ}$. 
  
In calculating of the intensity and the emission spectrum of the opaque tori,   
  we employed a standard ray-tracing formulation similar to that used in the calculation of relativistic lines
  from geometrically thin optically thick accretion disks \citep [e.g.][]{Cunningham1975}, 
  since it adequately takes into account effects such as gravitational redshift, lensing, kinetic time dilation and Doppler boosting. 
  Since only the propagation of radiation outside the torus is relevant, 
we set the emission and absorption coefficients to zero in the radiative transfer equation 
   along the rays emerging from the torus' boundary surface.    
We computed the relativistic frequency shifts at the surface boundary,  
  convolved this with a specified spatial profile for the source function of the emission 
  and obtained the emission spectrum.  

Figure~4 shows the profiles of emission lines from an opaque torus and a geometrically thin, optically thick disk 
  with the same inner and outer radius. 
The inner radius of the model accretion disk is much larger than 
  the minimum radius allowed, which is about $1~r_{g}$ for a maximally rotating black hole. 
A major difference between a torus and a thin disk is that 
  self-obscuration can occur for the torus at high viewing inclinations, 
  while the first-order emission from the upper disk surface is always visible by a distant observer.    
Thus, for an opaque torus, 
  self-eclipsing blocks the emission from the inner torus regions where relativistic effects are most severe 
  and the emission suffers the highest and lowest frequency shifts. 
For a thin disk, the inner disk regions with the highest and lowest frequency shifts contribute to the total emission spectra.  
The line profiles of the torus and the disk in Figure~4 show little  difference at low viewing inclinations. 
This is easily understood, because the torus and disk not only look similar when they are viewed pole-on, 
  they show similar radial dependences in the surface emissivity distribution 
  and the entire upper emission surface obscured.   
For viewing inclinations $ i \sim 80^{\circ}$ or higher,  
   the line from the torus is narrower than the line from the disk, 
   and in particular the edge of its blue wing is at a lower energy than that of the line from the disk, 
   because the inner torus region where the bluest emission originates is eclipsed by the front limb.  
 
\subsubsection{Emission from translucent pressure-supported tori}  

   
Optically thin and translucent tori do not have a sharply defined emission surface. 
All parts within the tori contribute to the emission, 
  with the contribution weighted according to matter concentration (density) 
  and to the local values of the thermodynamic variables (e.g. temperature) relevant to the radiation processes considered. 
The low optical depth across the torus permits the transmission of the emission from high-order lensed images,  
  therefore the emission spectrum of the torus 
  is the sum of the spectra from all the orders of lensed images weighted by the optical depth of the ray.    
Figure~5 shows the intensity images of an optically thin radiative pressure dominated torus for different viewing inclination angles, 
  and Figure~6 shows the intensity images of an optically thin radiative pressure dominated torus with a lower internal radiation pressure.
   
For the  $\beta = 1.235 \times 10^{-5}$ optically thin torus model, the value of $\beta$ was chosen so that the inner and outer radii of the torus very closely matched the inner and outer radii of the opaque tori and thin disks discussed in the previous section, allowing a comparison of the resulting images and emission line profiles. The second model with $\beta = 5 \times 10^{-5}$, i.e. a smaller internal radiation pressure, illustrates the dependence of the torus size on $\beta$.   
The dynamical parameters of the two tori are $n = 0.21$ and $r_{k} = 12~r_{g}$, 
  and the spin parameter of the central  black hole is  $a = 0.998$.    
The emissivity takes the form $j \propto \rho$, and the light-of-sight optical depth across the tori $\tau \ll 1$.  
As seen in the figures, changes in the pressure ratio parameter $\beta$ alter the aspect ratio of the torus, 
  which determines to which degree self-eclipsing would occur for a given viewing inclination angle. 
In spite of this, the general emission properties as shown in the images are qualitatively similar, 
   because of the low optical depth in the torus.
In both tori, the intensity of the emission is strongest at the interior of the torus, where the density is high. 
The rotation of the torus causes frequency shifts in the emission and Doppler boosting of the emission's intensity.   
These relativistic effects are more obvious for high viewing inclination angles.   
The emission from the approaching limb of the torus is amplified and appears to be significantly brighter than the emission from the receding limb of the torus. 

Figure~7 shows the profile of an emission line from a radiative pressure dominated optically thin torus viewed at different inclination angles (top panel). 
The lines are broad and have an asymmetric profile,  
  characteristic of line emission from relativistic accretion disks viewed at moderate inclination angles, $i \approx 45^{\circ} - 70^{\circ}$ 
  \citep[see e.g.][]{Cunningham1975, Fabian2000}. 
In contrast to the lines from relativistic accretion disks, 
  the profiles of lines from optically thin tori do not change significantly 
  when the viewing inclination angle changes from $45^{\circ}$ to $85^{\circ}$ (cf.\ the line profiles in Figure 4). 
The asymmetric broad profiles of the lines from optically thick relativistic disks 
   are due to the combination of a number of effects: Doppler boosting, Doppler velocity shift, gravitational time dilation and gravitational lensing. 
Emission from an optically thin torus does not depend on the projected area of an emission surface element, 
  as is the case of the optically thick accretion disks or optically thick accretion tori.    
While the emission from an optically thin torus is modified by Doppler boosting, Doppler velocity shifting and gravitational time dilation,  
  it is less affected by gravitational lensing and the area projection effect.     
The transparency of the torus to the emission could complicate the process of using the emission lines 
   to diagnose the dynamical properties of the torus. 
First of all, the emission of the high-order lensed images, which are unobscured, could severely contaminate 
  the emission from the direct image.  
Secondly, two emission lines can cross-contaminate each other in the spectrum,  
  and because of the large relativistic shifts one line could cause absorption of another line that originates from a different region within the torus. 
For an optically thin torus, multiple rays corresponding to different energy shifts can hit the same pixel in the image plane. 
This is forbidden for ray-tracing in the case of optically thick disks or tori, 
  where each ray originates from a single point on the emission boundary surface 
  of the disk or the torus 
  and each pixel in the disk or torus image corresponds to a uniquely defined relativistic energy shift.    
We show in Figure~7 (bottom panel) that    
   two distinct lines can be blended easily and may be taken as a single emission line with a complex relativistic line profile.   
    

\begin{figure}[htbp]
\begin{center} 
  \includegraphics[width=0.23\textwidth]{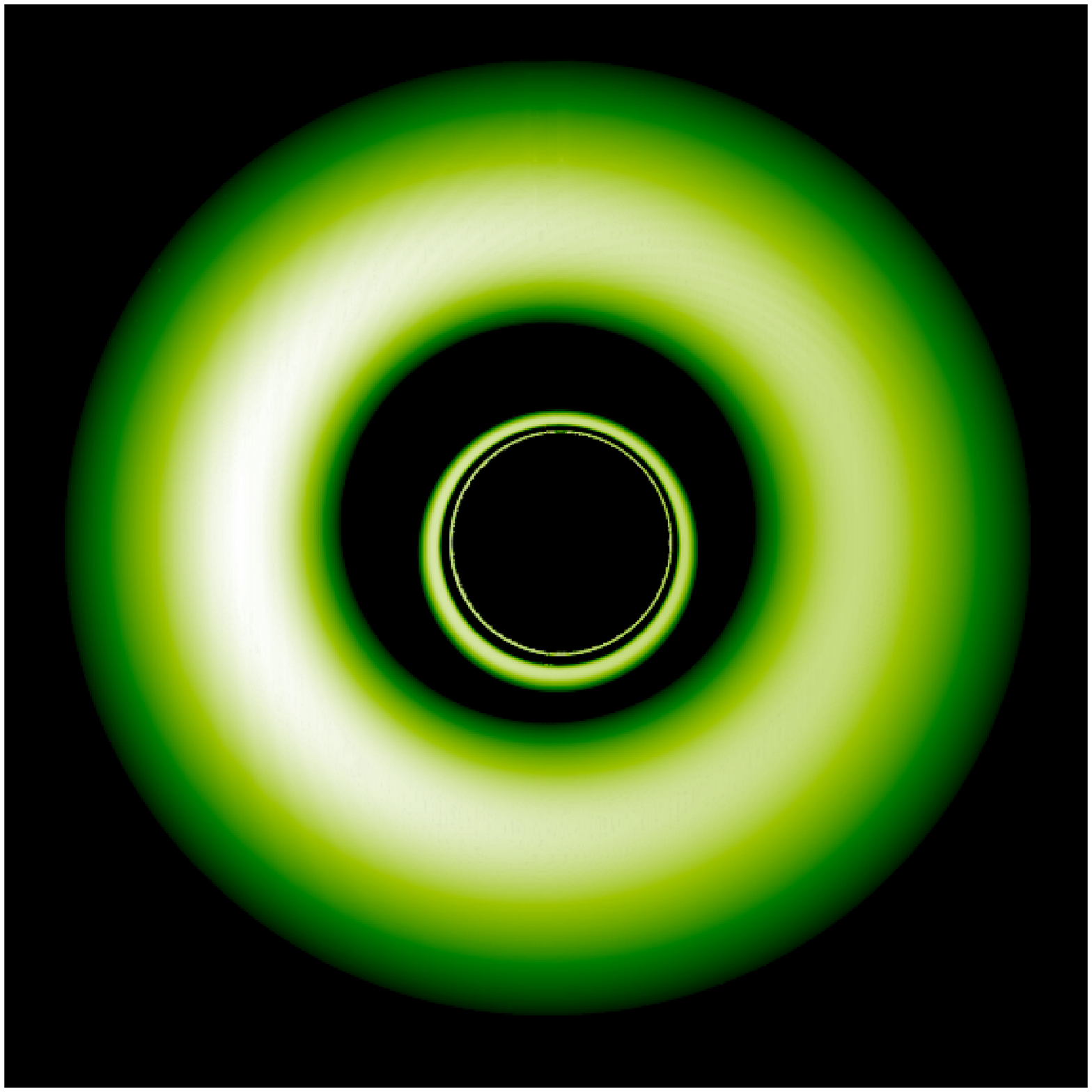}
  \includegraphics[width=0.23\textwidth]{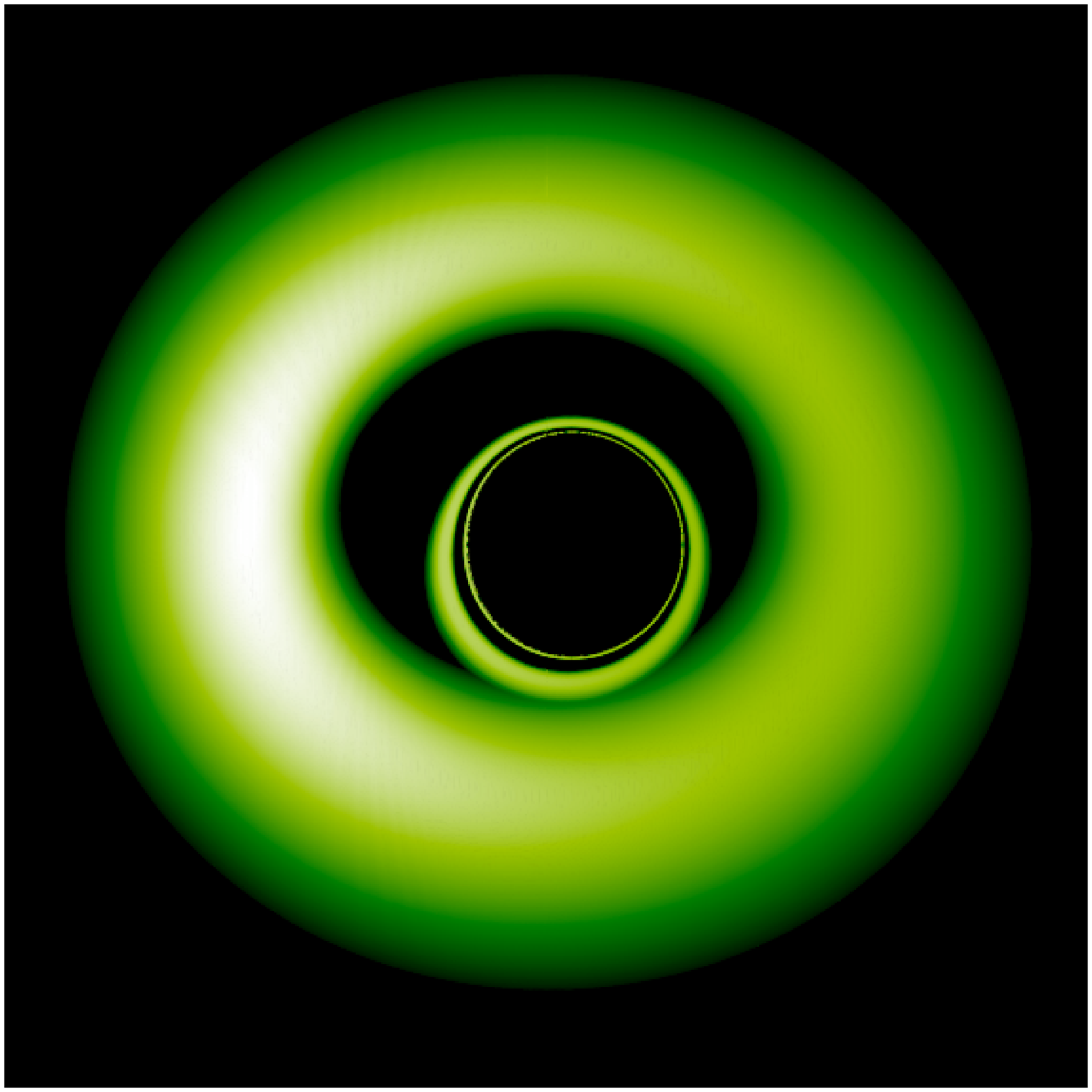}
  \includegraphics[width=0.23\textwidth]{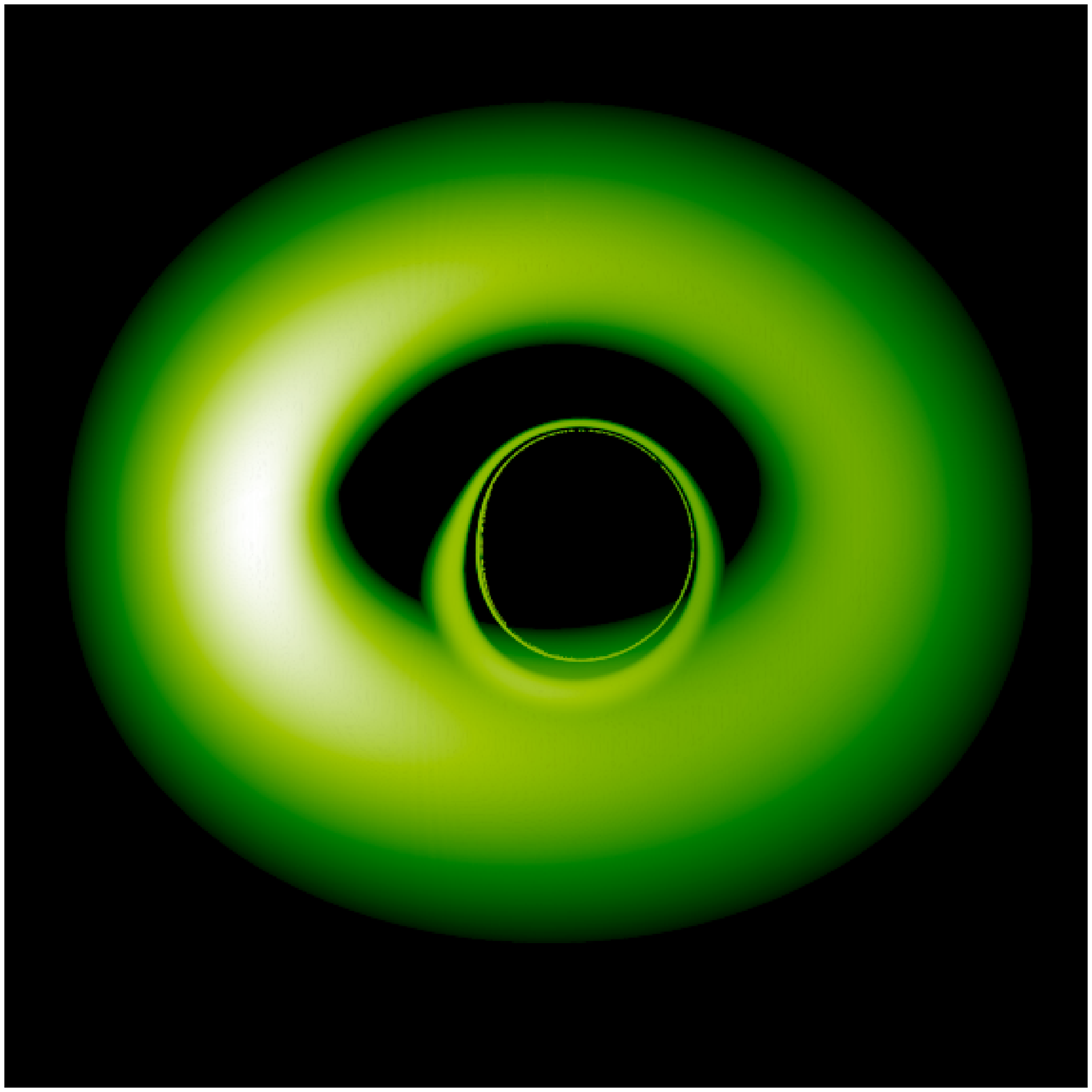}
  \includegraphics[width=0.23\textwidth]{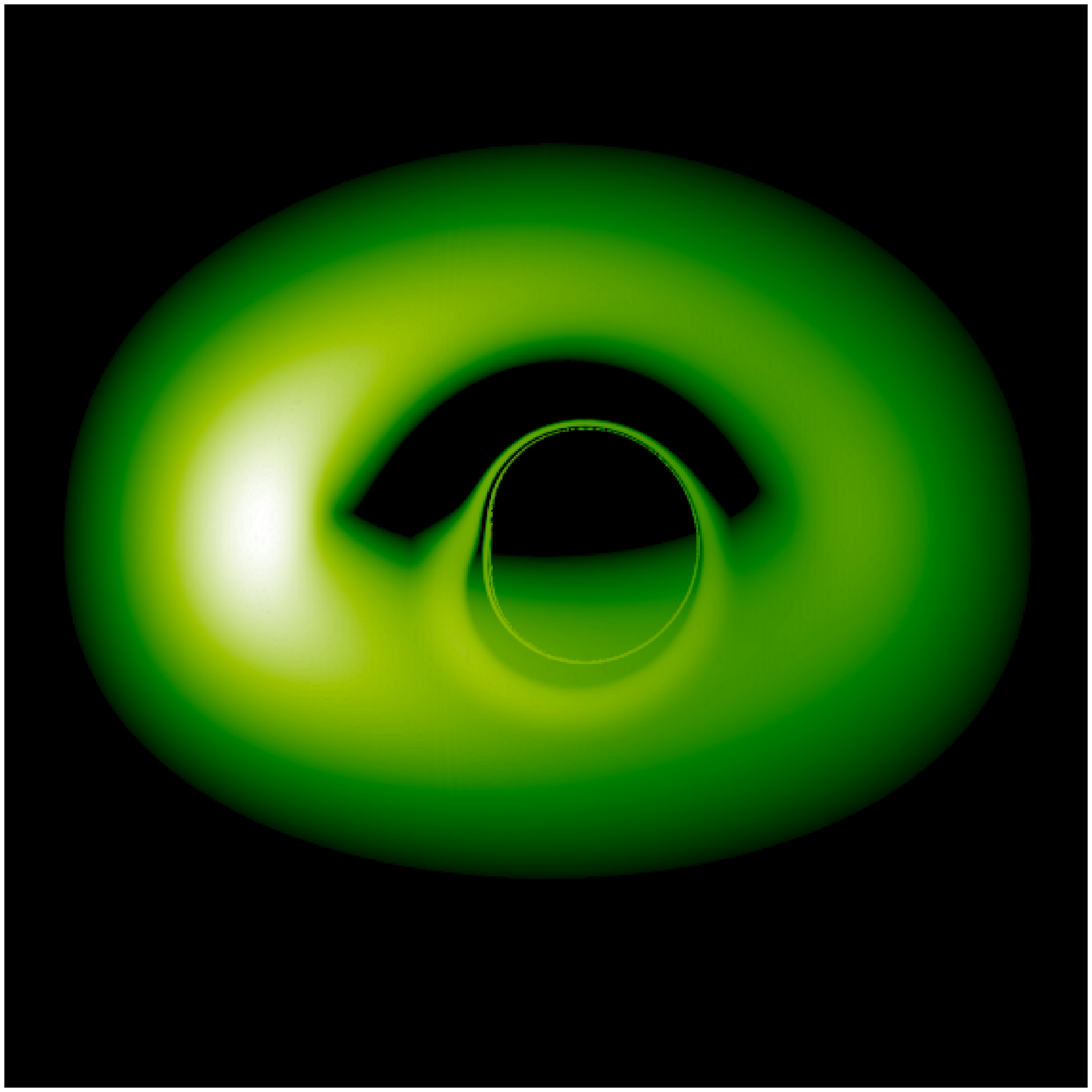}
  \includegraphics[width=0.23\textwidth]{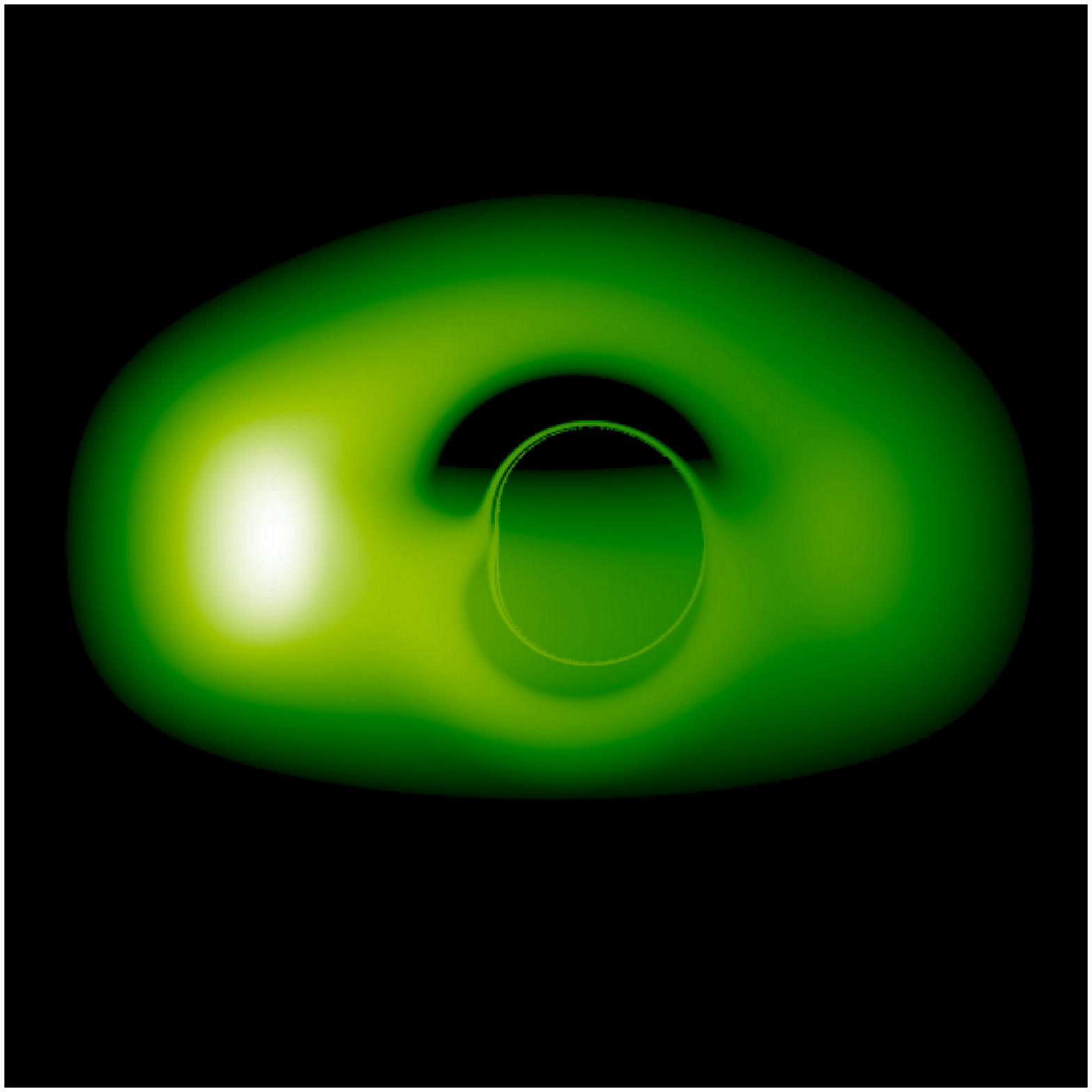}
  \includegraphics[width=0.23\textwidth]{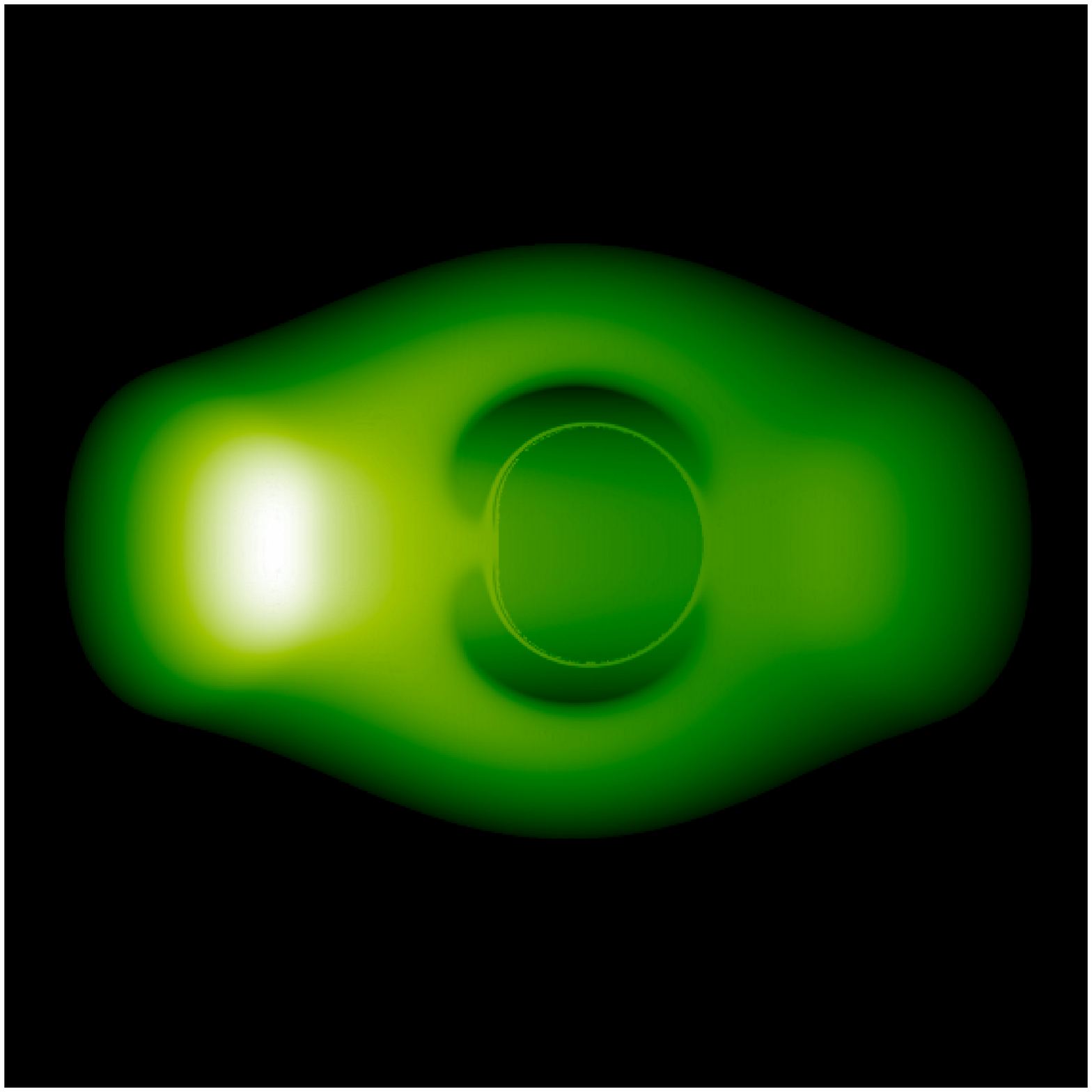} 
\caption{Surface brightness images of optically thin radiative pressure dominated accretion tori 
     viewed at inclination angles of $15^{\circ}$, $30^{\circ}$, $45^{\circ}$, $60^{\circ}$, $75^{\circ}$ and $89^{\circ}$ 
     (left to right, top to bottom).   
  The torus parameters are $n = 0.21$, $r_{k} = 12 r_{g}$ and $\beta = 1.235 \times 10^{-5}$.   
  The black-hole spin parameter $a=0.998$.   
 The brightness of each pixel represents the total intensity integrated over the entire spectrum. 
 The torus brightness is normalised such that the brightness of the brightest pixel in each image is the same.}
\end{center} 
\label{fig-5} 
\end{figure}  



\begin{figure}[htbp]
\begin{center} 
\includegraphics[width=0.23\textwidth]{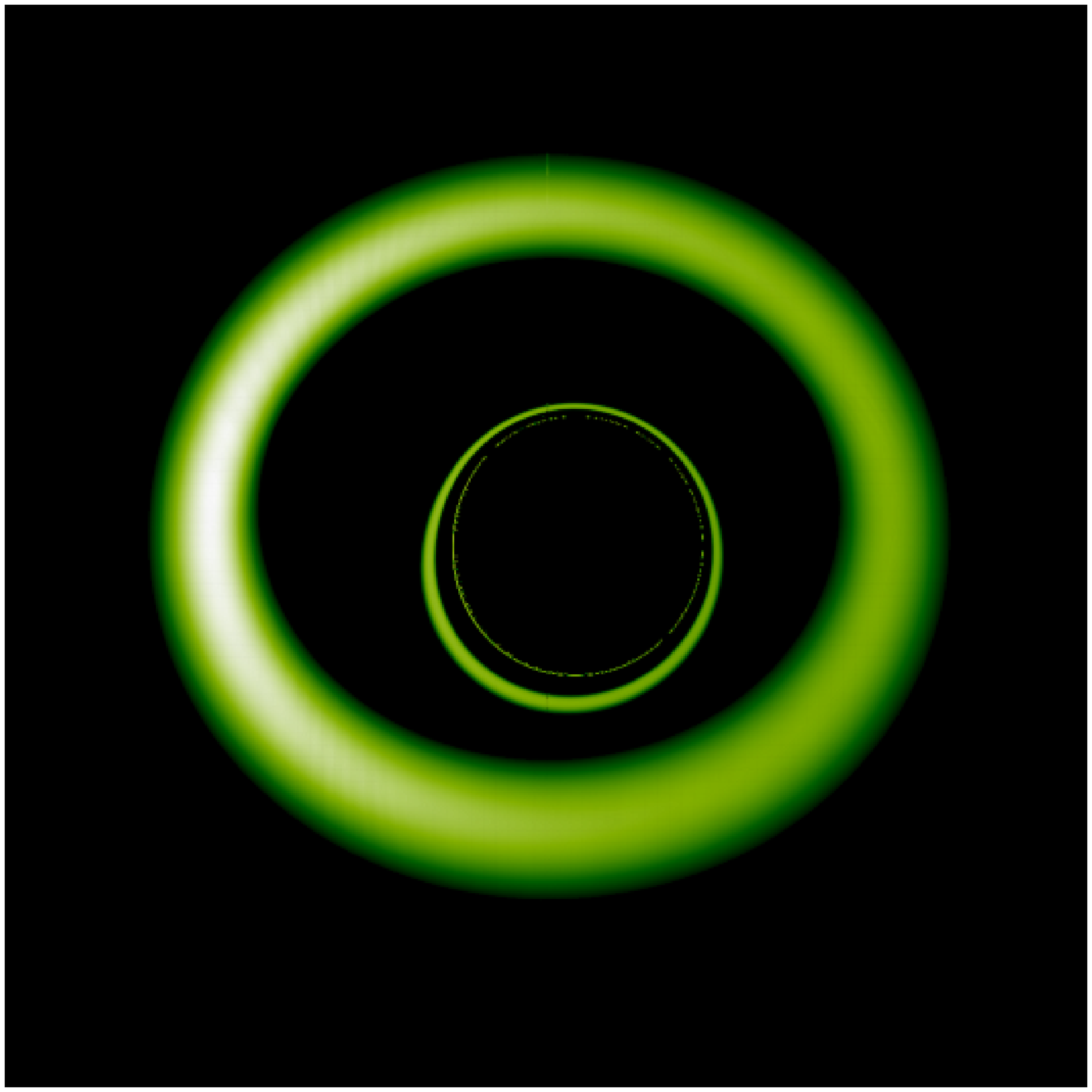}
\includegraphics[width=0.23\textwidth]{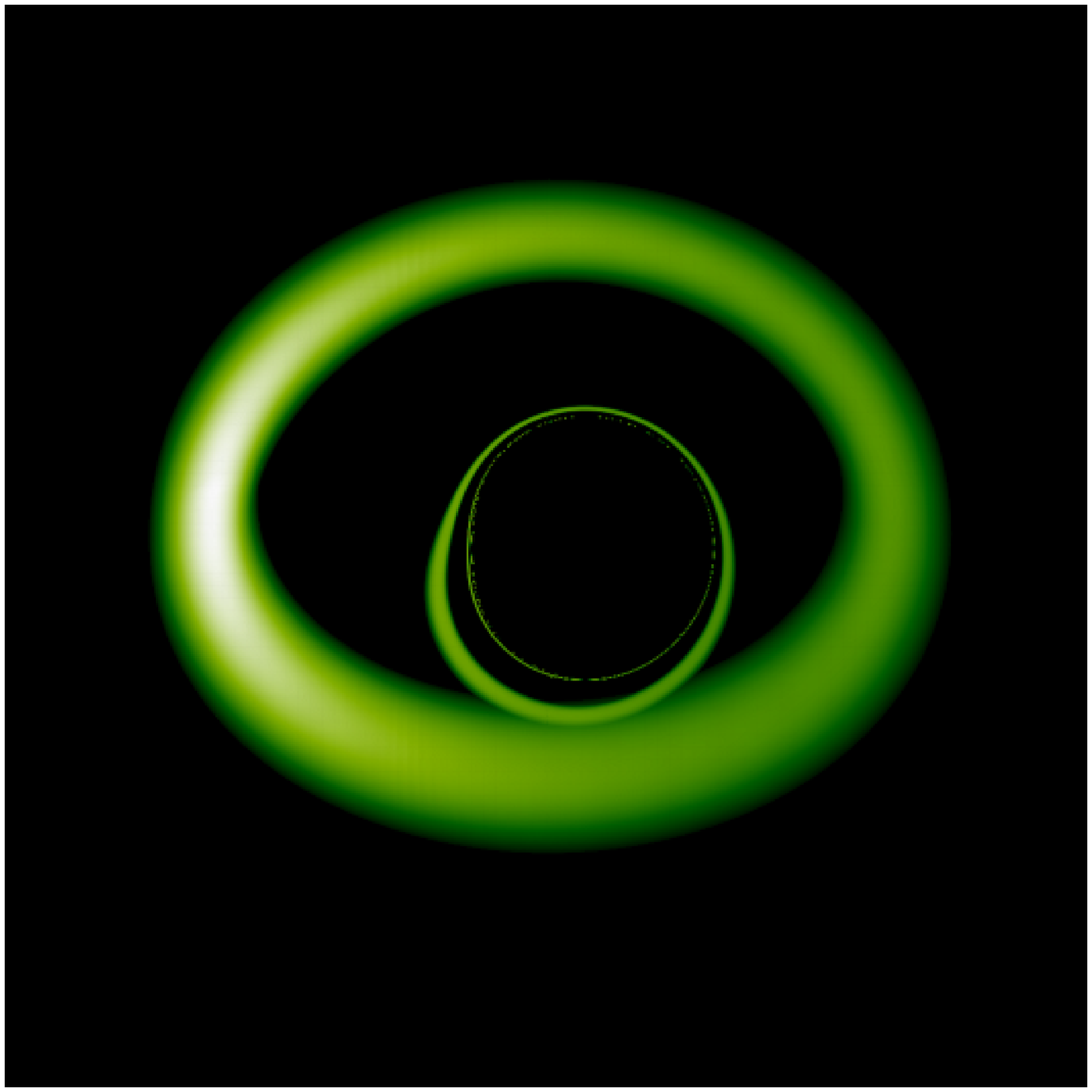}
\includegraphics[width=0.23\textwidth]{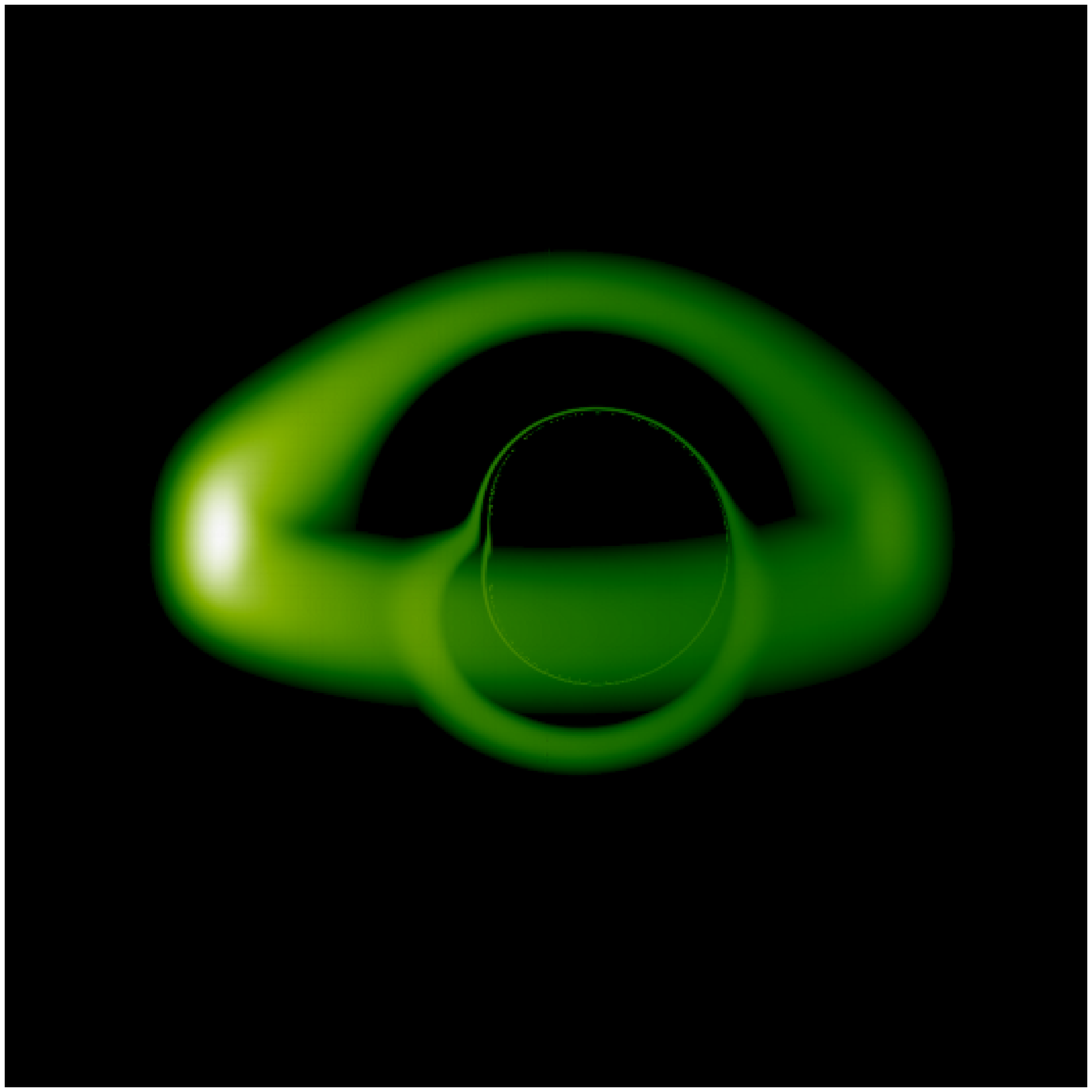}
\includegraphics[width=0.23\textwidth]{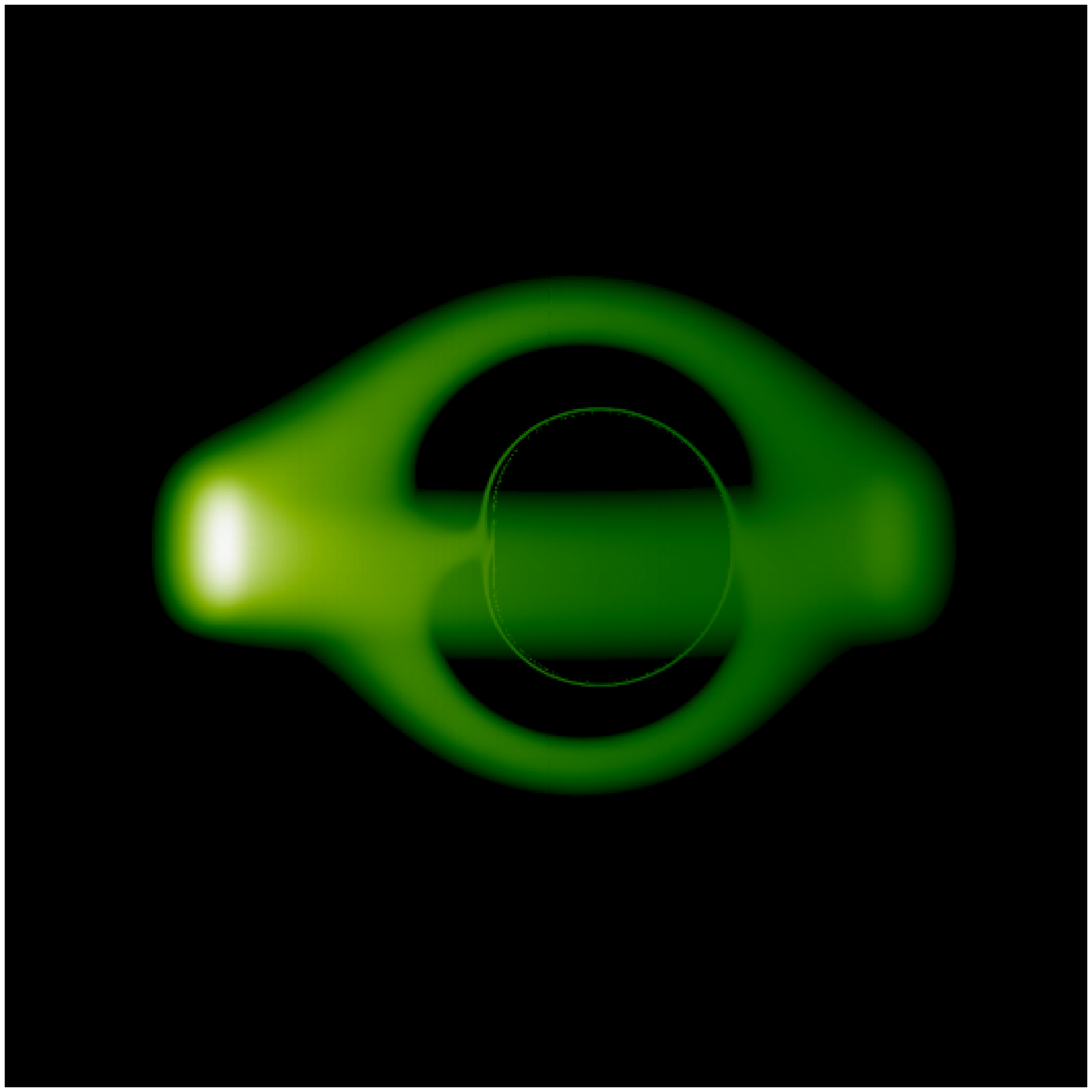}
\caption{Surface brightness images of optically thin radiation pressure dominated tori 
     viewed at inclination angles of $30^{\circ}$, $45^{\circ}$, $75^{\circ}$ and $85^{\circ}$ 
     (left to right, top to bottom).    
  The torus parameters are $n = 0.21$, $r_{k} = 12 r_{g}$ and $\beta = 5 \times 10^{-5}$.   
  The black-hole spin parameter $a=0.998$.   
 The brightness of each pixel represents the total intensity integrated over the entire spectrum. 
 The torus brightness is normalised such that the brightness of the brightest pixel in each image is the same.}
\end{center} 
\label{fig-6} 
\end{figure}  


  
\begin{figure}[htbp]
\begin{center} 
    \vspace*{0.2cm}  
   \includegraphics[width=0.425\textwidth]{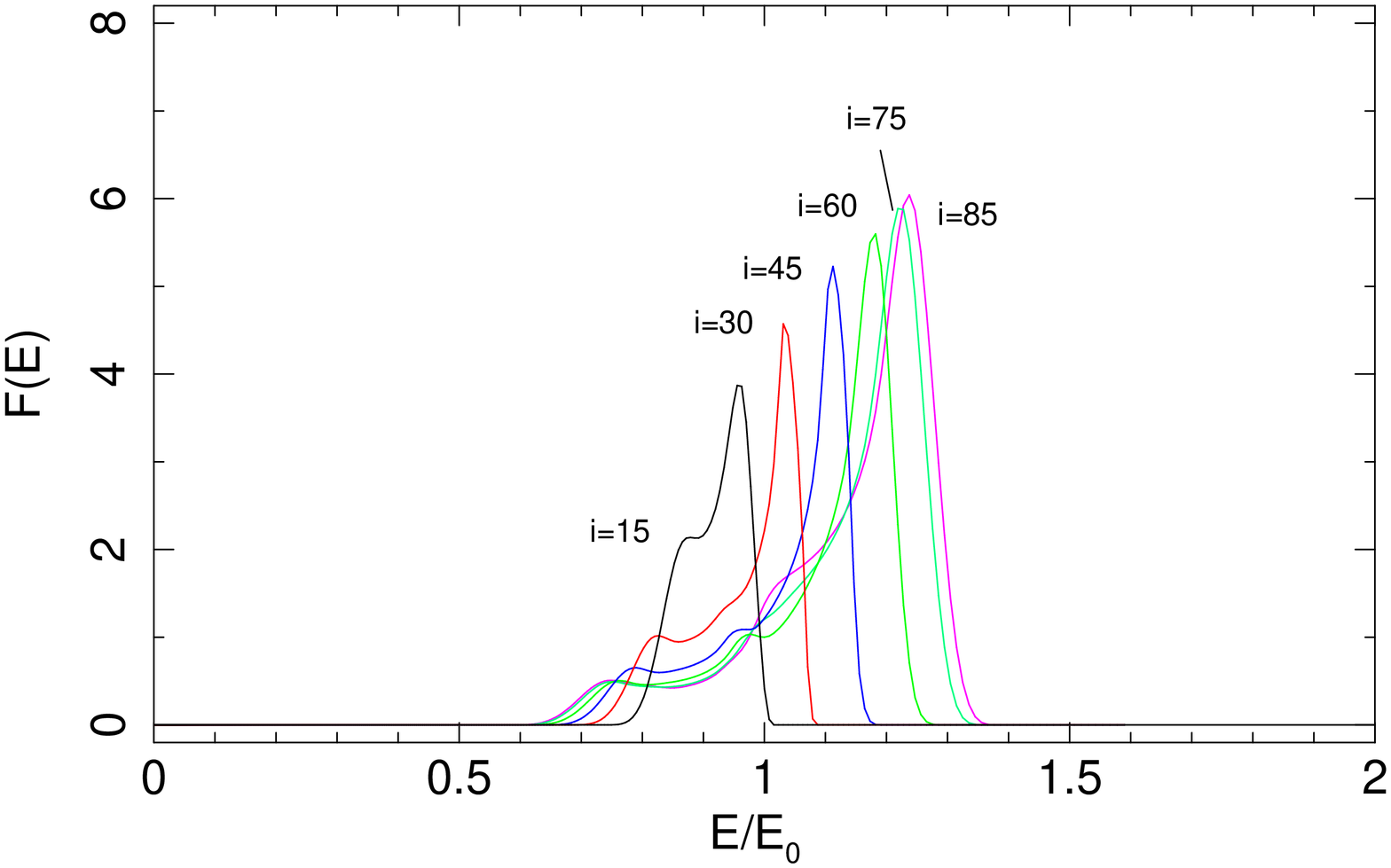}  \\ 
      \vspace*{0.12cm} 
   \includegraphics[width=0.425\textwidth]{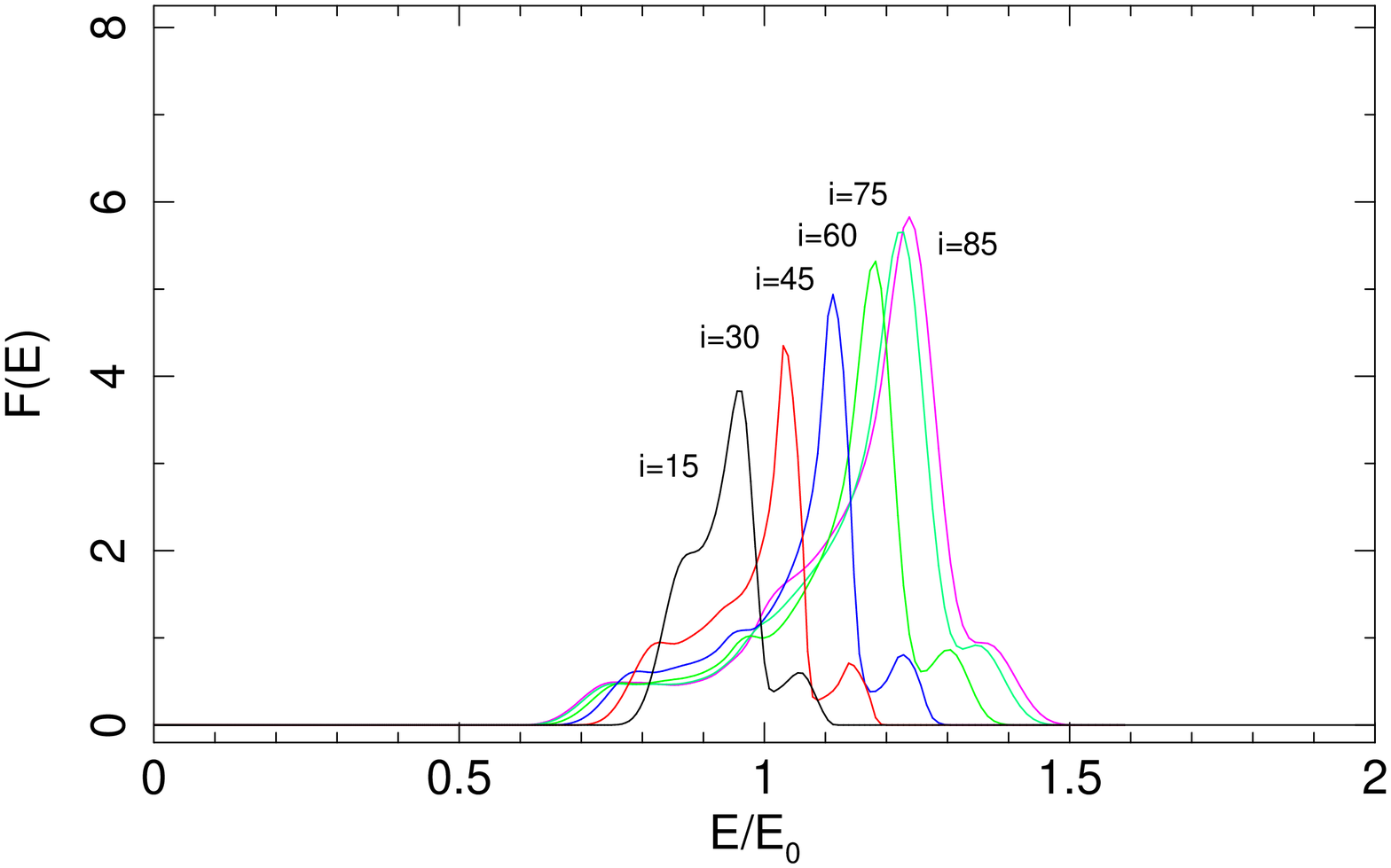}   \\ 
\caption{(Top panel) Profiles of an emission line from  an optically thin pressure supported structured accretion torus  
    viewed at different inclinations $i$.  
 The dynamical parameters of the torus are $n = 0.21$ and $r_{k} = 12\ r_{g}$.      
  The black-hole spin parameter $a = 0.998$.  
 The ratio of the gas thermal pressure to the total pressure $\beta = 1.235 \times 10^{-5}$. 
 The radius of the inner boundary $r_{in} = 8.528 r_{g}$ and the radius of the outer boundary $r_{out} = 20.246\ r_{g}$.  
 The line emissivity is proportional to the density $\rho$.  
 These line profiles are normalised such that the line flux $F(E_{0}) = 1$ when the torus  is viewed at $i = 60^{\circ}$.     
(Bottom panel) Profiles of composite profiles from two emission lines. 
  The torus and black hole parameters are the same as those for the lines in the top panel.    
   The line energies are such that one line has an energy 10\% higher than the other line 
        and the emissivity of the line with the higher line centre energy is 14\% of that of the line with the lower line centre energy 
        (cf.\ analogous to the relative properties of the Fe K$\alpha$ and K$\beta$ lines \citep{Holzer1997}).} 
\end{center}
\label{fig-7}
\end{figure} 


\subsubsection{Emission from quasi-opaque pressure-supported tori}  

The structured torus has density and temperature stratifications (see Figure~2)
  that give rise to opacity variation across the torus and opacity gradients along the line-of-sight 
  for different radiative processes.     
When the density is sufficiently high, the torus becomes opaque to radiation.    
Quasi-opaque structured tori have more complex emission properties than 
  their optically thin counterparts and the idealised optically thick rotationally supported tori in the previous sections. 
   
We investigated the emission properties 
  using our covariant radiative transfer formulation, conducting a full radiative transfer calculation for a model-structured torus.   
We considered two opacity sources, whose specific emissivities in the rest frame are given by 
\begin{eqnarray} 
  j_{0,1}(E_{0})  & =  & \mathcal{K} \  \bigg(\frac{n_{e}}{\mathrm{cm}^{-3}} \bigg)^{2}  \bigg(\frac{E_{0}}{\mathrm{keV}}\bigg)^{-1}\bigg(\frac{\Theta}{\mathrm{keV}}\bigg)^{-1/2} \mathrm{e}^{-E_{0}/  \Theta} ,   \\ 
  j_{0,2}(E_{0})  & = & C \ \bigg(\frac{n_{e}}{\mathrm{cm}^{-3}}\bigg) \ \! \bigg(\frac{E_{0}}{\mathrm{keV}}\bigg)^{-2.5} , 
\end{eqnarray}  
  where $\Theta = kT$ is the relativistic temperature and $\mathcal{K}=8\times 10^{-46}\ \!  \mathrm{erg} \  \mathrm{s}^{-1} \mathrm{cm}^{-3} \ \! \mathrm{Hz}^{-1}$. $C$ is a normalisation, dependent on $\beta$, chosen such that both $j_{0,1}$ and $j_{0,2}$ are equal at $E_{0}=0.1$ keV, at the torus centre. For $\beta = 1.235 \times 10^{-5}$, $C = 2.162\times 10^{-45}\ \!  \mathrm{erg} \  \mathrm{s}^{-1} \mathrm{cm}^{-3} \ \! \mathrm{Hz}^{-1}$, whereas for $\beta = 5 \times 10^{-5}$, $C=2.681\times 10^{-45} \ \!  \mathrm{erg} \  \mathrm{s}^{-1} \mathrm{cm}^{-3} \ \! \mathrm{Hz}^{-1}$. The electron number density is defined as $n_{e}=\rho/\mu m_{H}$.
The corresponding specific absorption coefficients are 
\begin{eqnarray} 
  \alpha_{0,1}(E_{0}) & = & B_{1}\  \bigg(\frac{n_{e}}{\mathrm{cm}^{-3}}\bigg)^{2}\  \sigma_{T} f_{1}(E_{0}) \ \mathrm{cm}^{-1} \ ,  \\ 
  \alpha_{0,2}(E_{0}) & = & B_{2}\  \bigg(\frac{n_{e}}{\mathrm{cm}^{-3}}\bigg)\  \sigma_{T} f_{2}(E_{0}) \ \mathrm{cm}^{-1}  \   ,   
\end{eqnarray} 
  where $\sigma_{T}$ is the Thomson cross-section, and 
  $f_{1}(E_{0})$ and $f_{2}(E_{0})$ are functions of photon energy. 
The torus is optically thin to the first process, 
  but is partially opaque to the second process.  
We therefore set $B_{1} = 0$. 
Without loss of generality
  we consider an energy independent absorption, which implies $f_{2}(E_{0})=1$.  
The normalisation constant $B_{2} $ 
  is chosen such that $\alpha_{0,2} r_{out} \sim 1 - 5$ across the torus, 
  where $r_{out}$ is the outer boundary radius of the torus.
Note that the first process has a  similar density and temperature dependence to thermal free-free emission. 
The second process mimics a free-electron scattering-like process 
  that converts photons with different energies indiscriminately into a power-law energy distribution.   
      

\begin{figure}[htbp]
\begin{center} 
  \includegraphics[width=0.23\textwidth]{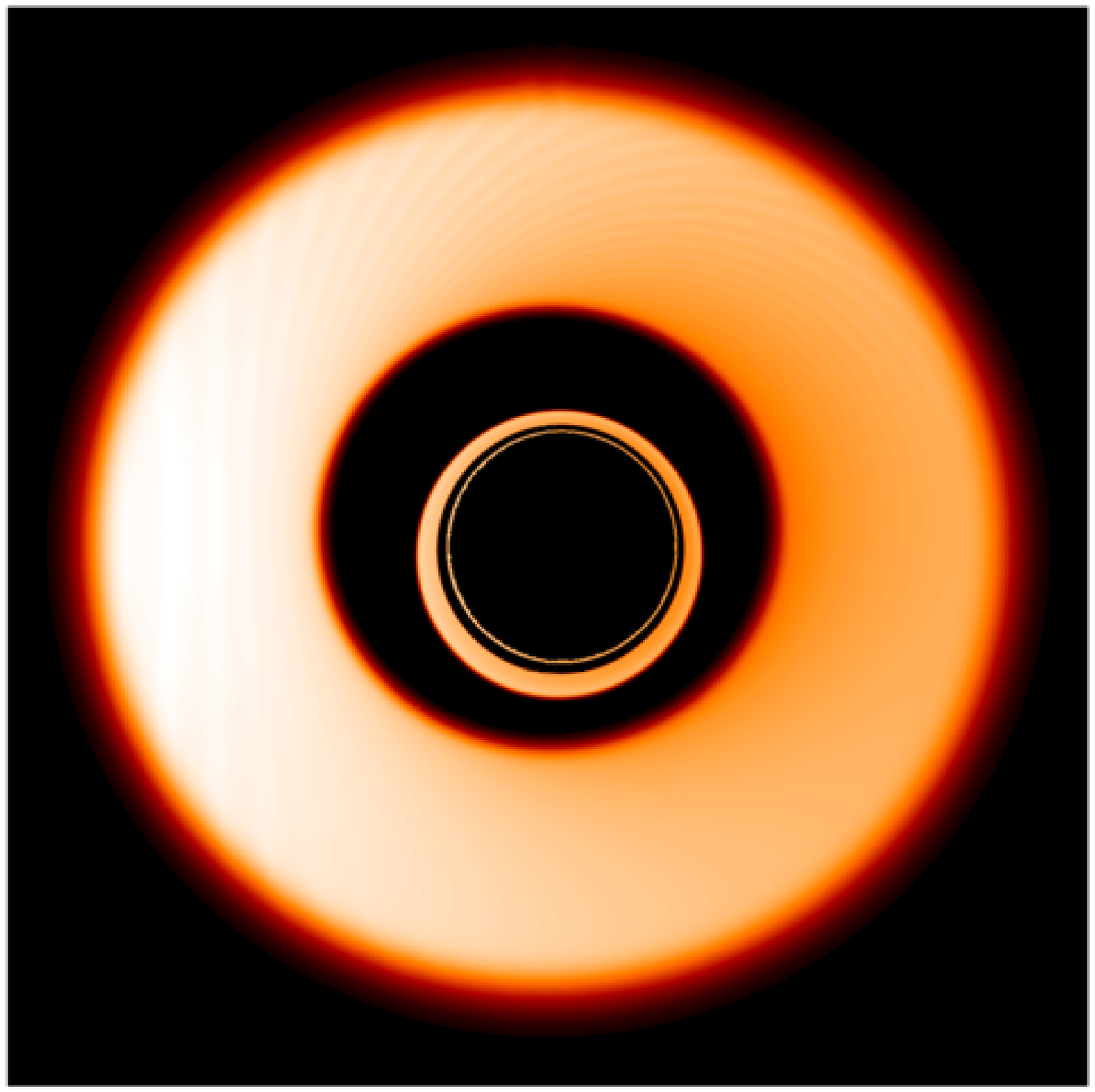}
  \includegraphics[width=0.23\textwidth]{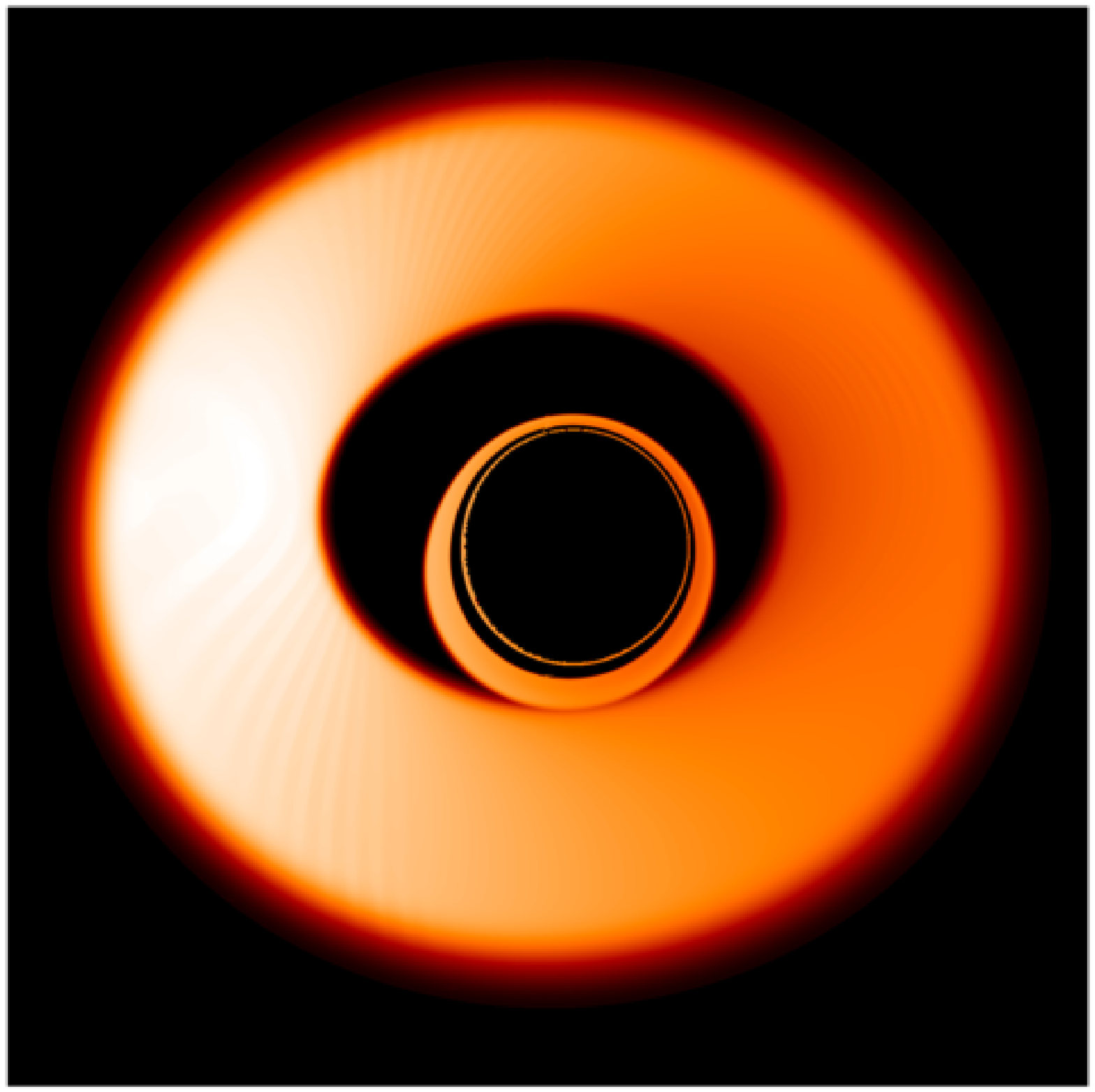}
  \includegraphics[width=0.23\textwidth]{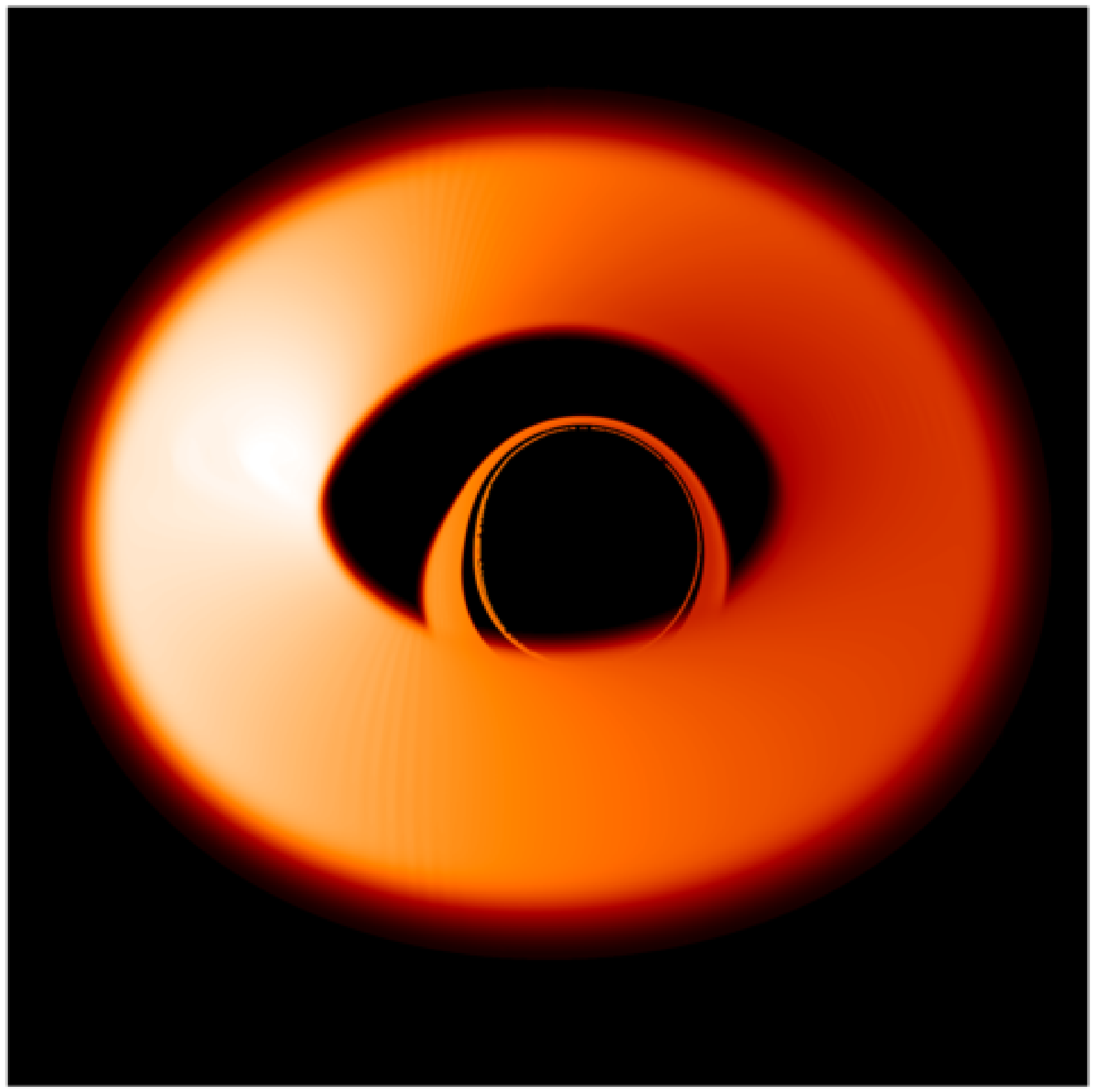}
  \includegraphics[width=0.23\textwidth]{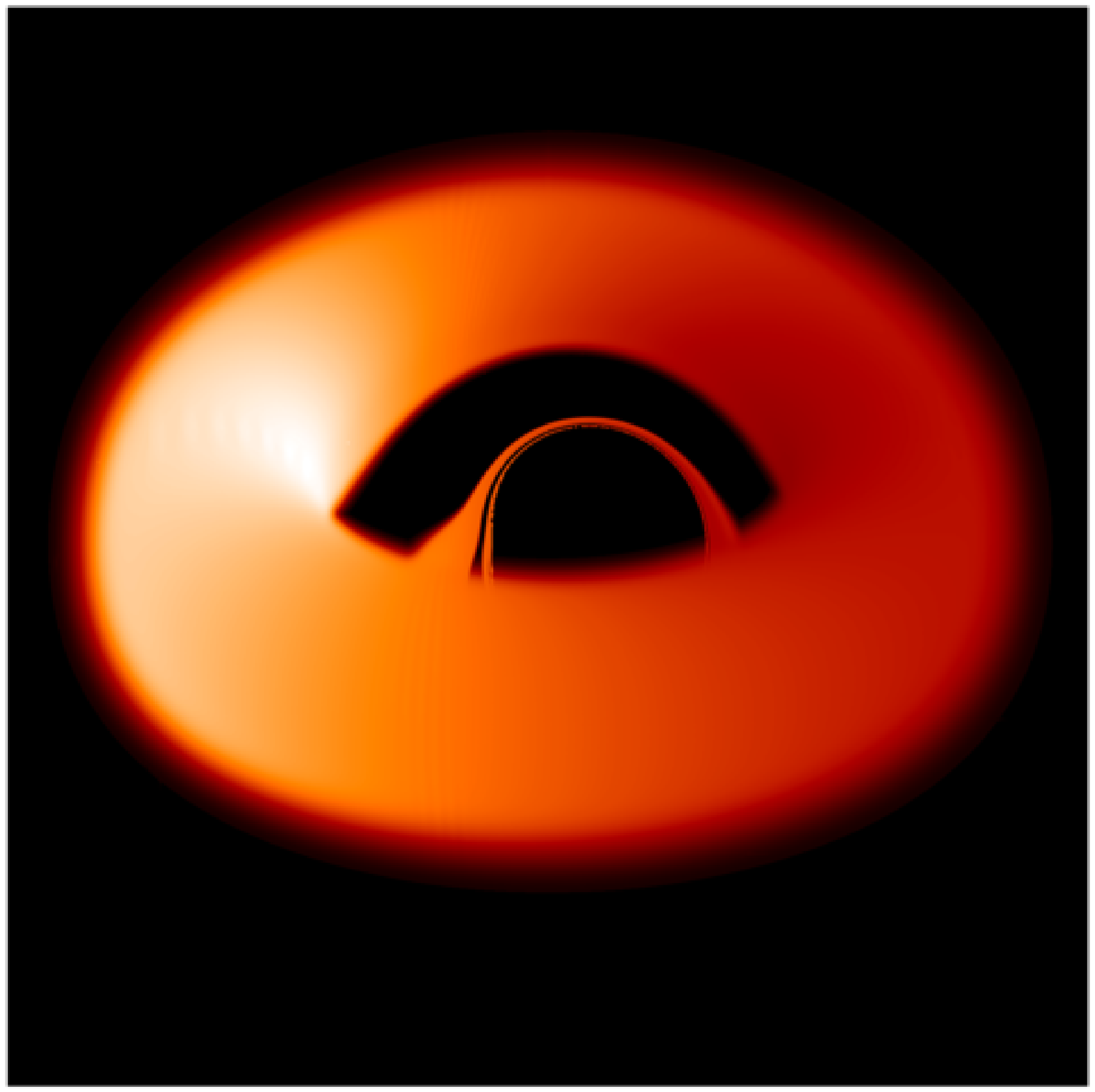}
  \includegraphics[width=0.23\textwidth]{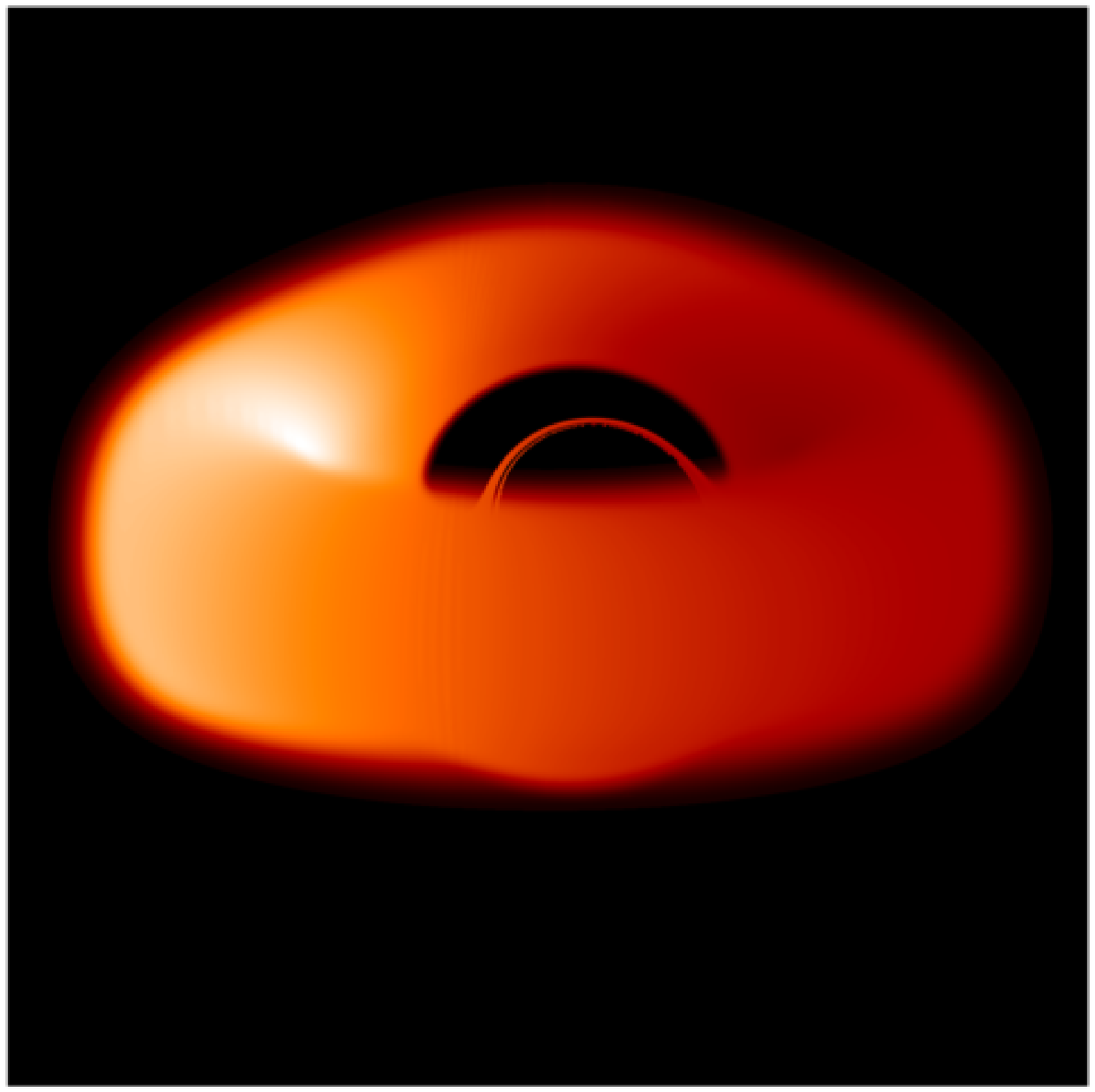}
  \includegraphics[width=0.23\textwidth]{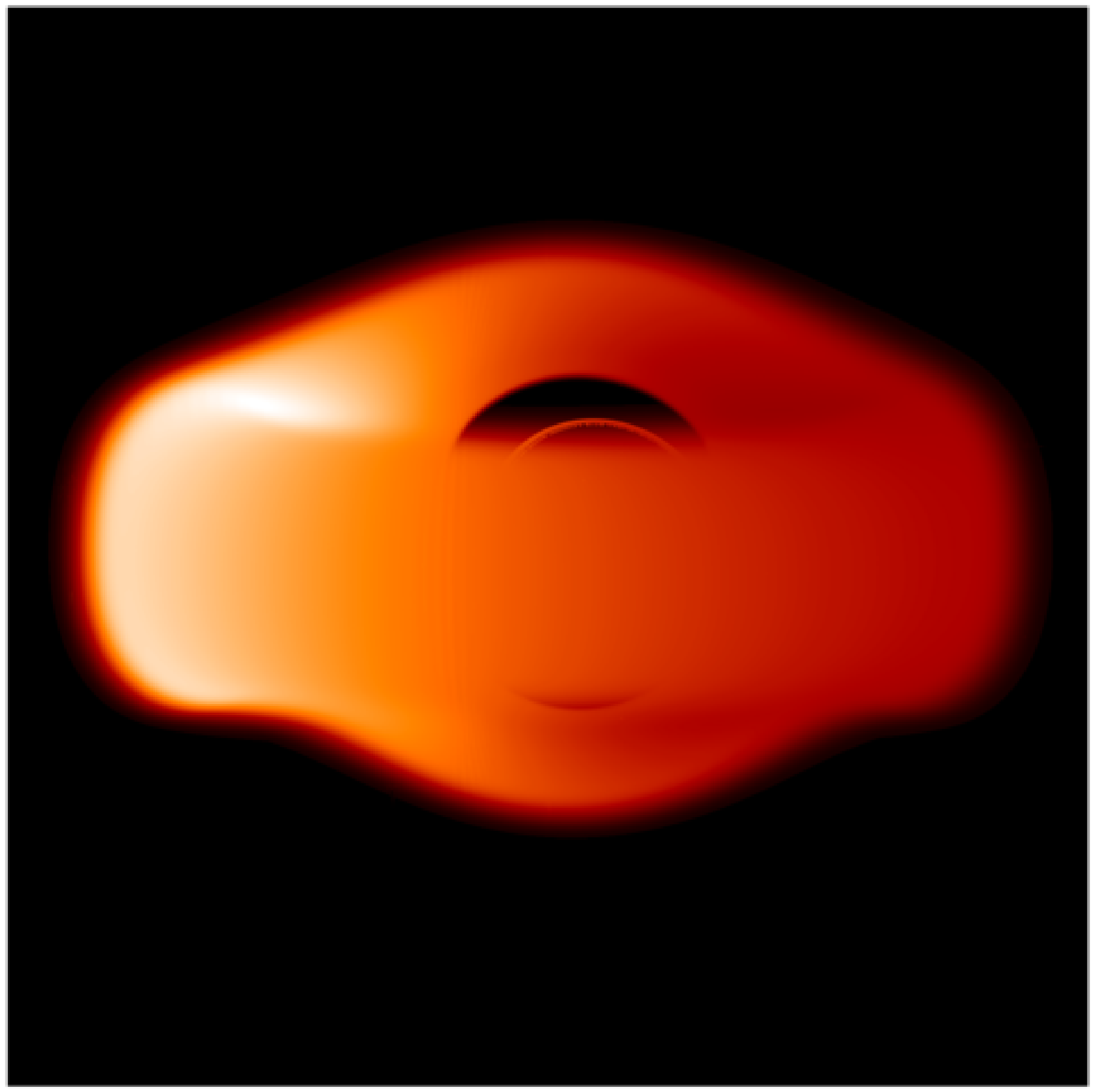}  \\ 
\caption{Surface brightness images of opaque radiative pressure dominated accretion tori 
     viewed at inclination angles of $15^{\circ}$, $30^{\circ}$, $45^{\circ}$, $60^{\circ}$, $75^{\circ}$ and $85^{\circ}$ 
     (left to right, top to bottom).   
  The torus parameters are $n = 0.21$, $r_{k} = 12 r_{g}$ and $\beta = 1.235 \times 10^{-5}$.   
  The black-hole spin parameter $a=0.998$.     
  The emissivity is provided by emission from two spectral lines, and the two opacity sources given in equations (59) and (60). Absorption is provided through the Thomson cross-section in equation (62).
  The brightness of each pixel represents the total intensity integrated over the entire spectrum. 
  The torus brightness is normalised such that the brightness of the brightest pixel in each image is the same.}
\end{center} 
\label{fig-8} 
\end{figure}  



\begin{figure}[htbp]
\begin{center}
\includegraphics[width=0.23\textwidth]{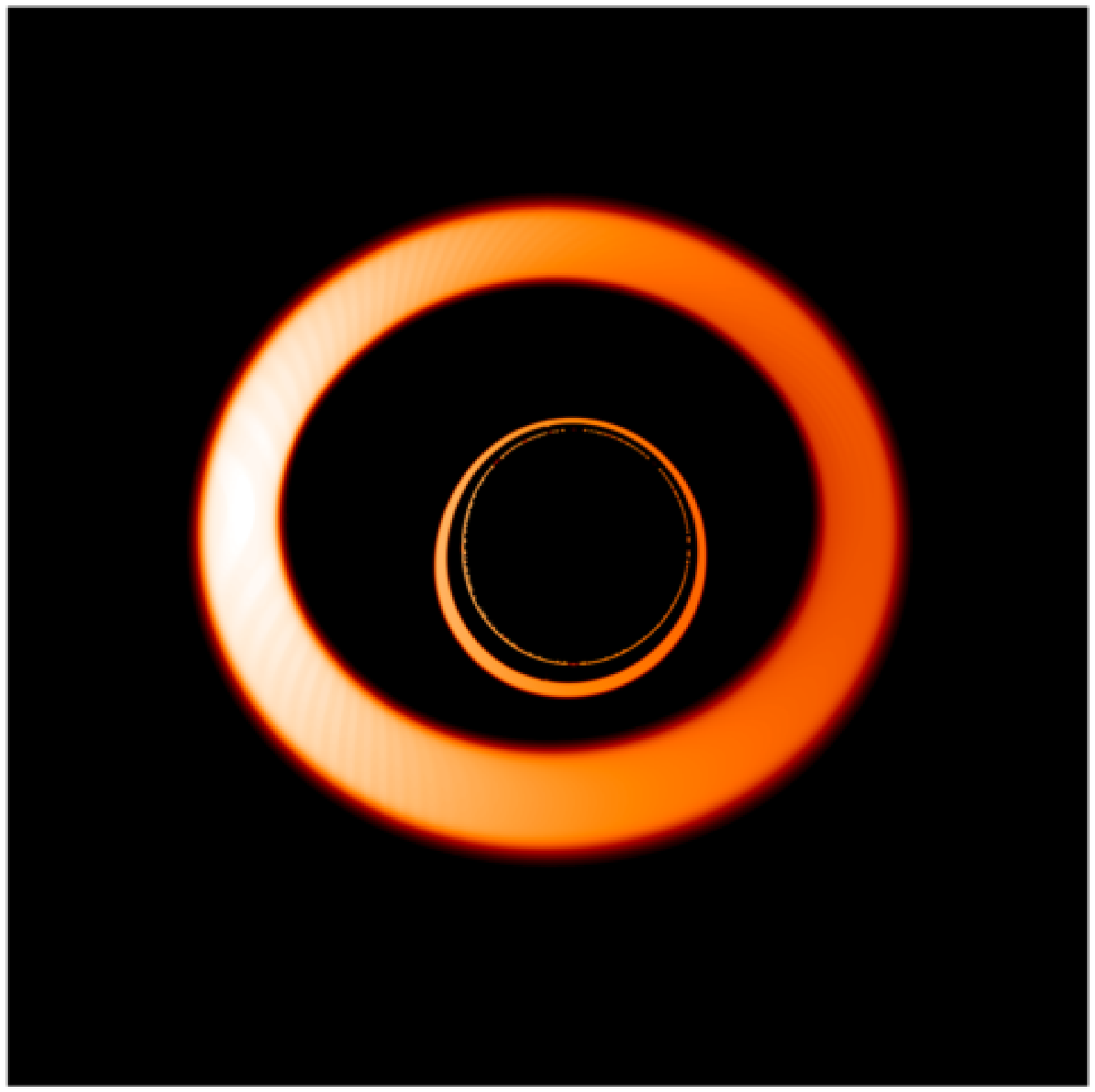}
\includegraphics[width=0.23\textwidth]{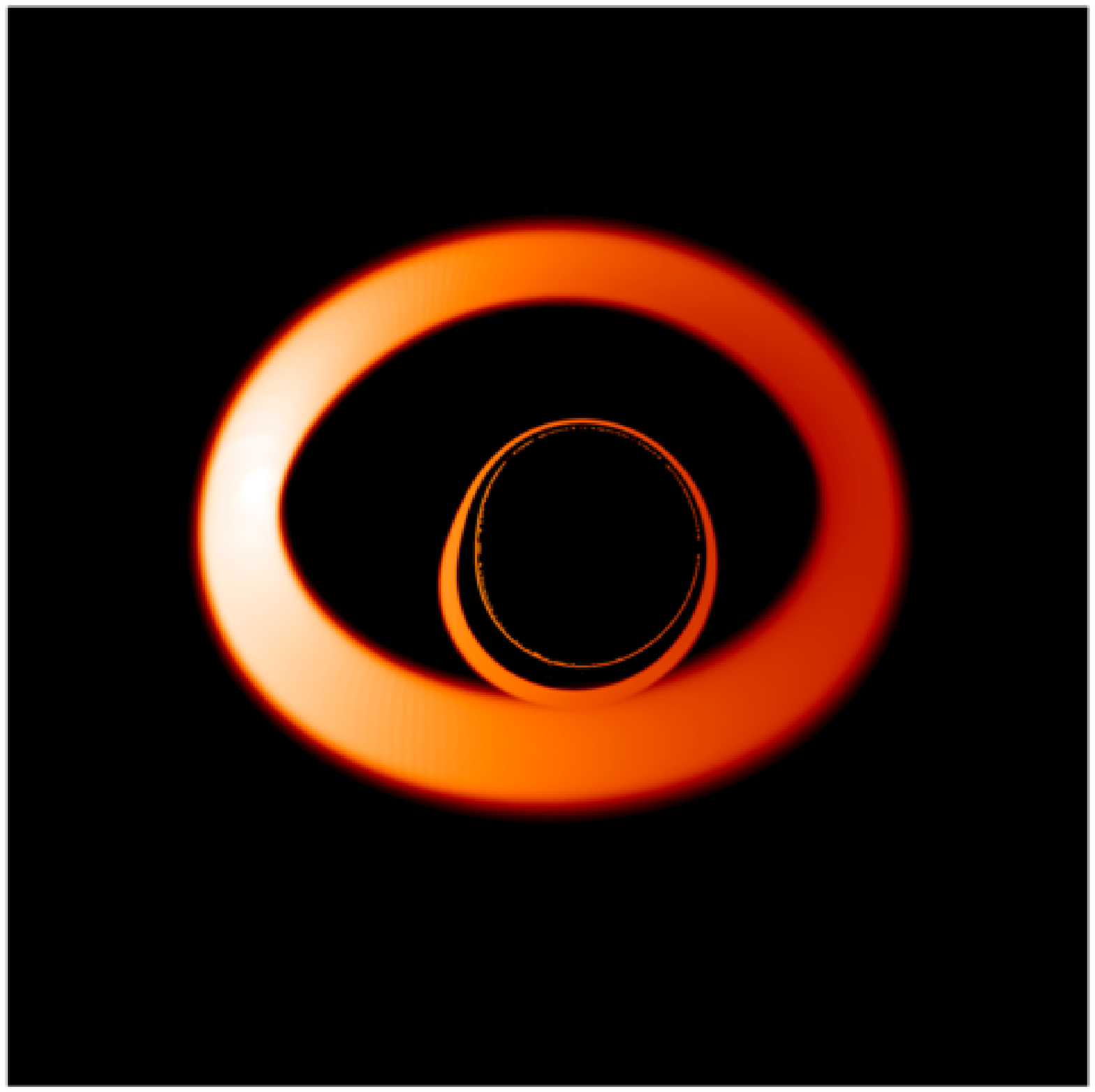}
\includegraphics[width=0.23\textwidth]{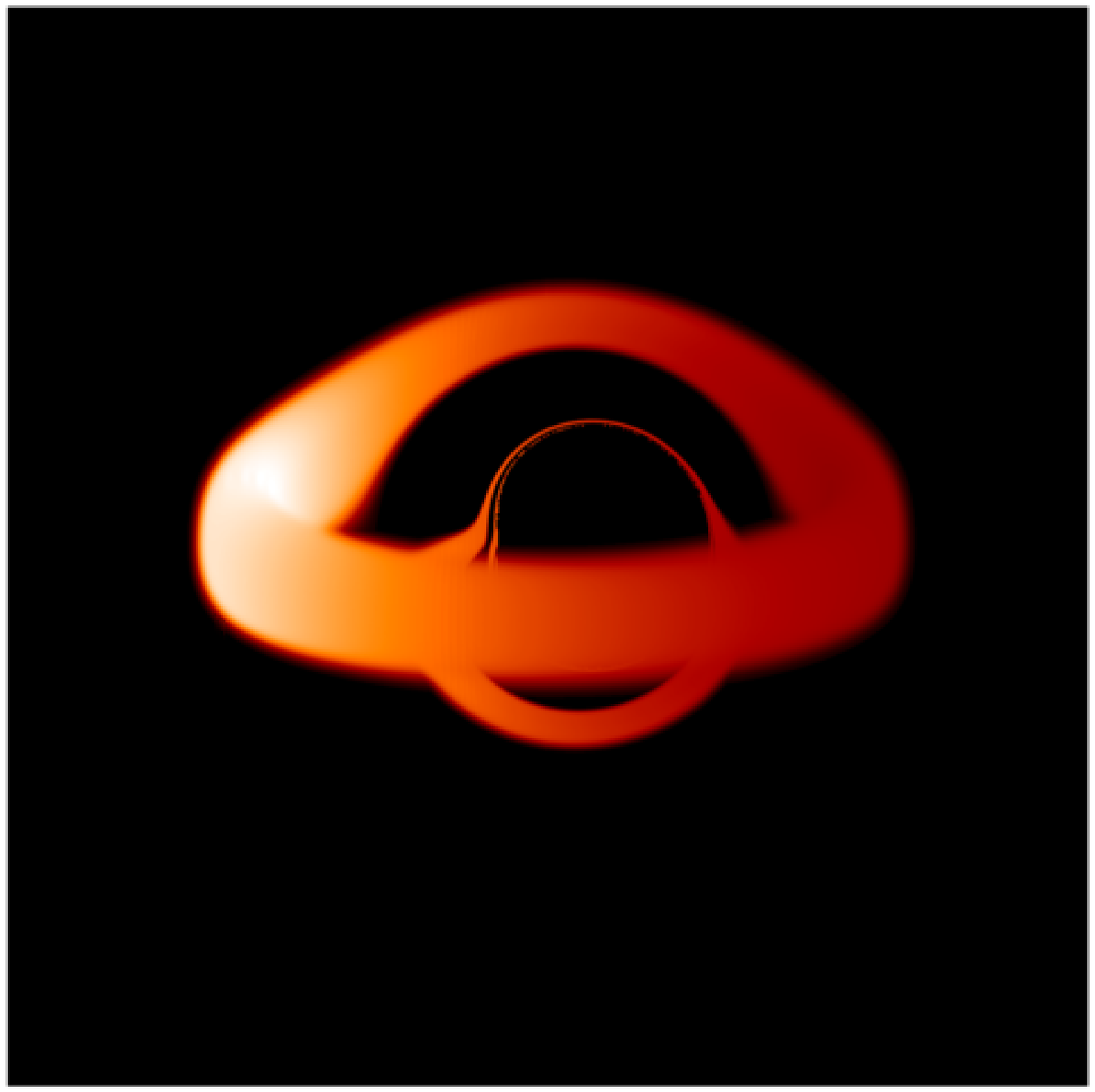}
\includegraphics[width=0.23\textwidth]{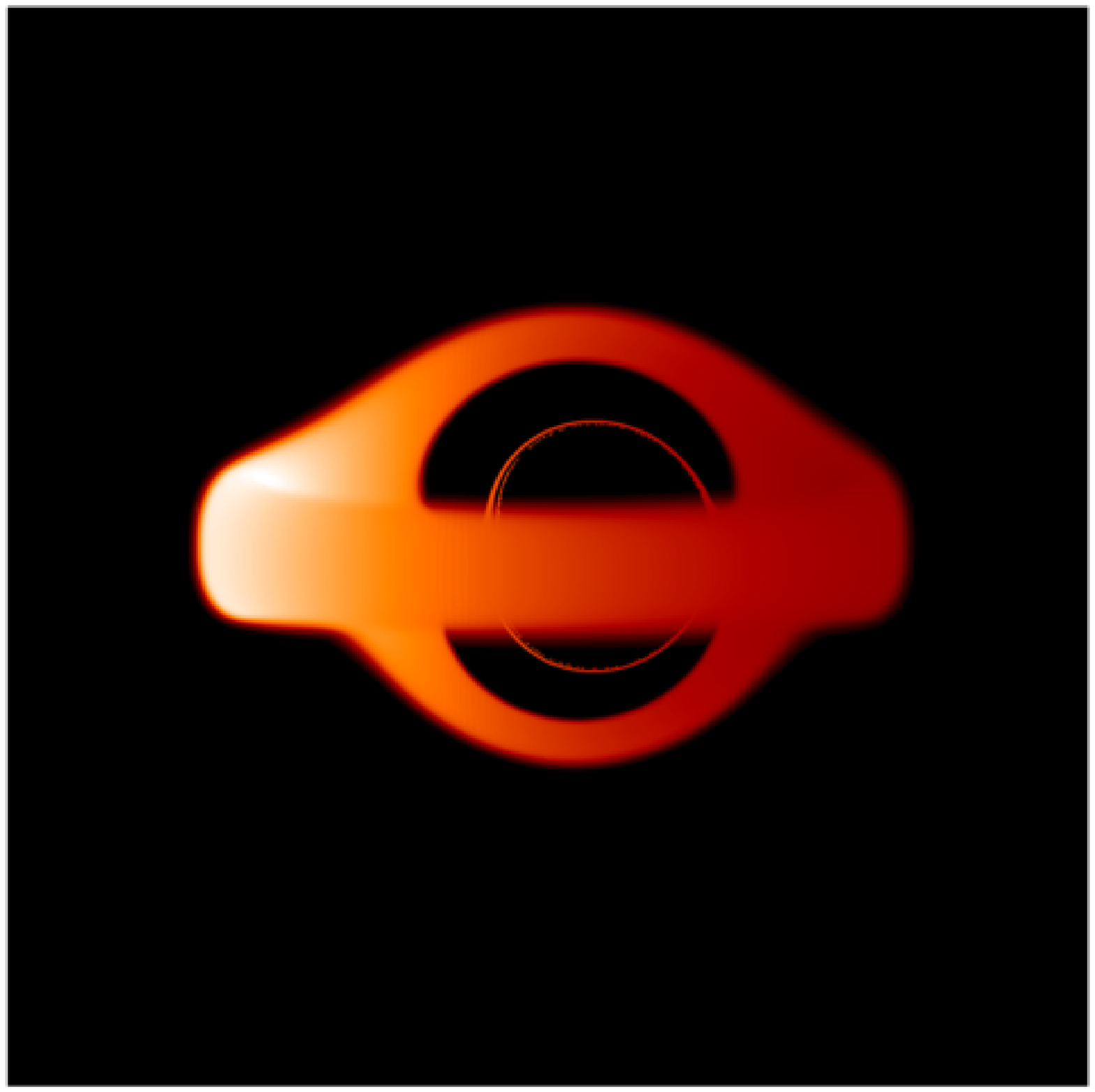}
\caption{Surface brightness images of quasi-opaque accretion tori around extreme Kerr (a=0.998) black holes, 
   viewed from left to right, top to bottom, 
    at observer inclination angles of $30^{\circ}$, $45^{\circ}$, $75^{\circ}$ and $85^{\circ}$. Same parameters as figure 8, but now $\beta = 5 \times 10^{-5}$.
    Line emission and continuum emission are included. 
   These tori also self-occult and higher-order emission is greatly suppressed at high inclination angles. 
   The images are normalised such that the brightest pixel in each image is of the same intensity.}
\end{center}
\end{figure}



Figures~8 and 9 show intensity images of opaque radiative pressure dominated tori using the aforementioned spectral parameters. These tori differ from the optically thin case in that the emissivity and opacity have changed, causing significant differences in the intensity images. Towards the outer-edges of the tori there is now obvious limb darkening (dark red).
There is now an additional temperature dependence on the emission that is caused by the temperature stratification of the torus, nearly tangential rays sample only the cooler surface layers of the torus. However, rays almost perpendicular to the torus surface have a much higher probability of traversing the hotter layers deeper beneath the torus surface. 
Consequently, surfaces viewed face-on by an observer will appear to be much brighter than those viewed at higher inclination angles. For these tori we have assumed $\tau \gg 1$.

\subsection{Some remarks} 

\subsubsection{Optically thin vs optically thick emission}  

For the optically thin tori in figures 5 and 6, the low optical depth means that these tori are almost transparent, allowing the entire torus volume to contribute to the emission. The emission from the approaching side of the torus is much stronger due to Doppler boosting, which causes an increase in the projected velocity of the gas along the line of sight towards the observer. The receding limb's projected velocity is decreased and consequently dimmed. At higher viewing inclination angles geodesics travel through more emissive material than at lower inclinations. Since the fast-moving gas in the inner region of the torus is visible, blueshifting and beaming are very strong and emission from the approaching side of the torus dominates these images. The overall brightness of the optically thin tori at high inclination angles is much brighter than at lower inclinations, because at high inclinations geodesics may sample significantly more emissive material in the limbs.
\\ \\
For the rotationally supported opaque tori, emission comes from the surface of the torus only. Since the opaque torus models are optically thick, the front limb of the torus obscures the fast-moving inner regions of these tori at higher inclination angles. This is reflected in their emission spectra in figure 4, where for $i\ge45^{\circ}$ the observed flux decreases rapidly. For the optically thin torus model, emission is observed from the entire volume of the torus, not just the surface. Since the fast-moving gas from the inner-region contributes to the observed emission line profile (Figure 7), this gives rise to the more pronounced red and blue wings, which are observed even at high inclination angles. Additionally, for optically thin tori the observed line profiles appear to be roughly monotonically increasing with observer inclination angle. This is in contrast to the line profiles from optically thick accretion disks and tori in Figure 4, which decrease in amplitude and appear wedge-shaped for inclination angles beyond $45^{\circ}$ (\cite{Kojima1991}, \cite{Beckwith2_2004}, \cite {Fuerst2004}). As was found in \cite{Fuerst2006}, altering the black hole spin does not affect the observed torus images or emission line profiles significantly. The inner edges of the tori presented in this paper do not extend as close to the event horizon as thin disks. Consequently, the red wing of their line profiles is not as extended. The shape of these tori, as well as the location of their inner edges, depends much more on the distribution of pressure forces within the tori themselves. If these models are realistic, it would prove difficult to derive the central black hole spin from the observed spectral information.
\\ \\
The emission from quasi-opaque tori is restricted to a thin surface skin layer 
    that is similar to cases of the opaque accretion tori and disks. However, the 
    quasi-opaque torus is distinguishable from the fully opaque accretion tori and accretion 
    disks by the effects arising from gradients in the optical depth. This effect is particularly 
    noticeable near the edges of the quasi-opaque tori, where significant limb darkening 
    occurs. These limb effects are a manifestation of the temperature and density stratification  
    within the tori. When looking at the edge of a torus, for a given optical depth, cooler,
    less dense material is observed than when looking towards the
    torus centre. Just above the surface of the quasi-opaque torus is an optically thin photosphere. As can be seen in figures 8 and 9, photons traversing only the edges of the limb of the torus are weakly attenuated by the cool material and so appear significantly darker than the rest of the torus surface, i.e. only line photons are observed. Hotter neighbouring regions appear to be much brighter because continuum photons are attenuated by the local emission and absorption properties of the emitting medium.
\\ \\
We illustrate the limb effect with a simple phenomenological model of an emitting and absorbing medium, in local thermodynamic equilibrium (LTE), at a uniform temperature \textit{T}, with a background source of intensity $\mathcal{I}(0)$. In the absence of scattering, taking $\tau_{0}=0$ at the outer edge of the limb, equation (18) implies
\begin{equation}
\mathcal{I}(\tau_{\nu})=\mathcal{I}(0)\mathrm{e}^{-\tau_{\nu}}+\mathcal{B}(\mathit{T})(1-\mathrm{e}^{-\tau_{\nu}}), \nonumber
\end{equation}
where $\mathcal{B}(\mathit{T})$ is the Planck function. For low optical depths, $\tau_{\nu}\rightarrow 0$ and so $\mathcal{I}(\tau_{\nu})\rightarrow \mathcal{I}(0)$, i.e. the observed intensity of the limb tends towards the background intensity, is independent of temperature, and emission from regions deeper within the torus becomes negligible. In our calculations we assumed $\mathcal{I}(0)\ll 1$ and so the limb darkening effect is obvious, and in strong contrast to the rest of the torus image.

\subsubsection{Gravitationally induced line resonance in 3D flows} 

Consider a two-level atomic system. Absorption of a photon causes the excitation of an electron from a lower energy state to a higher energy state. Conversely, a photon is emitted when an electron is de-excited from a higher energy state to a lower energy state.   
The energy of the absorbed and the emitted photon is the same as the energy difference between the two states. 
Such a system can be considered an oscillator.   
Under general conditions, two oscillators with different intrinsic frequencies at different locations would not be resonantly coupled, and two lines from different atomic transitions would not exhibit resonant behaviour without coupling to either other lines or an optical pump.   

However, in the vicinity of black holes, where relativistic effects are severe and gravity is extreme,  
  lines from different atomic transitions can couple and exhibit resonance phenomena.  
As shown in numerous publications  
  \citep[e.g.][]{Cunningham1975, Fabian1989, Stella1990, Fanton1997, Reynolds1999, Fabian2000, Fuerst2004}, 
  the gravitational frequency shifts of lines from relativistic accretion disks around black holes can be severe.  
Photons propagating upwards, out of the gravitational well of the black hole, are subject to energy redshifts, 
  whereas photons propagating in the opposite direction, deeper into the gravitational well, are blueshifted.


\begin{figure}[htbp]
\begin{center}
\includegraphics[width=0.4\textwidth]{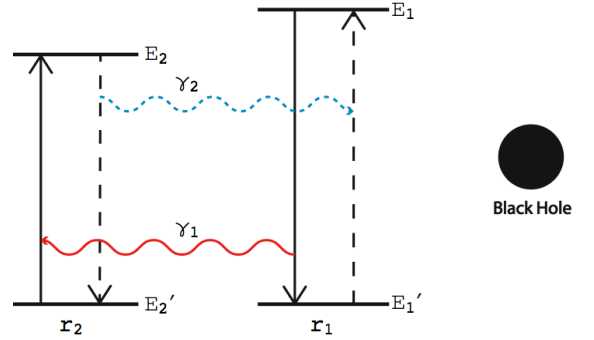}
\caption{Schematic illustration of the resonance between two lines with $E_{1}$ and $E_{2}$, 
      emitted at radial distances $r_{1}$ and $r_{2}$ from a black hole, 
      where $r_{2} > r_{1}$. 
   In the rest frames $E_{1}>E_{2}$. 
   }
\end{center}
\end{figure}
   

Consider a radiative transition process occurring in a medium at a radial distance $r_{1}$ from the black hole,  
  which consequently emits a photon of energy $E_{1}$. 
The photon propagates outward, and its energy is redshifted. 
When the photon reaches a distance $r_{2}$, its energy becomes $E_{2}$, lower than $E_{1}$,  the energy at its point of emission. 
Suppose that the photon encounters an electron and is absorbed. 
The electron is excited to a higher energy state. 
The electron is then de-excited and returns to its original state, emitting a photon of energy $E_{2}$. 
This photon, however, propagates inward. 
When it reaches $r_{1}$, its energy has been gravitationally blueshifted and is now $E_{1}$. 
Again suppose that this photon encounters an electron, is absorbed, 
  and causes the electron to be excited to a higher energy level. 
This electron is subsequently de-excited, and again emits a photon of energy $E_{1}$. 
The photon propagates outward once more, is absorbed by an electron, and causes another excitation. 
The later de-excitation of this electron leads to the emission of another photon of energy $E_{2}$, 
  and the photon propagates inward again... 
This process would persist, forming a resonant feedback cycle between the transitions of the two lines 
 with different rest-frame energies  
 (see Figure~10).  

Such a line resonance occurs easily in 3D flows 
  but not in geometrically thin accretion disks or 2D flows.  
This can be understood as follows.  
In the Schwarzschild space time, 
  for a photon with an energy $E_{1}$ located at $r_{1}$, 
  one can also find a closed surface corresponding to an energy redshift of $\Delta E$, 
  such that the photon energy  becomes $E_{2} = E_{1} - \Delta E$, 
  located at $r_{2} > r_{1}$. 
The converse also holds, and a photon of energy $E_{2}$ located at $r_{2}$ 
  always finds a closed surface corresponding to a blueshift of $\Delta E$, 
  such that the photon energy increases to $E_{1} = E_{2} +\Delta E$. 
A photon emitted from the vicinity of a black hole must pass through 
  the closed surface with a specific energy shift before reaching infinity, 
  and in 3D radial flows the entire surface is in principle embedded completely in the flow. 
A 2D flow cannot completely contain such a surface 
  and hence the photon can escape to infinity, 
  passing through the surface with a specific energy shift at a location not inside the flow.   
The situation is similar in the Kerr space-time. 
A detailed quantitative analysis of such a line resonance, 
  and relativistic radiative transfer calculations including these masing effects warrants further investigation. We leave this to a future article.

\section{Conclusions}

We have derived the radiative transfer equation from first principles, 
  conserving both particle number and phase-space density. 
The equation is thus manifestly covariant, and  
  it is applicable in arbitrary 3D geometrical settings 
  and for any pre-defined spacetime metric. 
The more general form of the equation explicitly considers 
a covariant particle flux with mass. 
We have found that 
  ind addition to modifying the geodesic trajectories that connect the observer and the emitting regions,  
  the presence of particle mass in the radiative transfer equation 
  introduces a mass-dependent aberration, which reduces the intensity gradient along the ray. 
In the zero-mass limit 
  this general radiative transfer equation recovers its original form for the massless particles.

We carried out demonstrative numerical general relativistic radiative transfer calculations 
  with a ray-tracing algorithm constructed from the formulation. 
Different 3D accretion tori around rotating black holes 
  with different geometrical aspects, physical structures, emission properties, 
  and optical depth variations were considered.    
We demonstrated that radiative transfer calculations based on the formulation 
  are able to deal with the complexity in the various combinations and convolution 
  of relativistic, geometrical, physical and optical effects.  
Our calculations clearly showed the significant role that  
  structures and optical depth, and their gradients, 
  together with geometrical and relativistic factors, 
  play in shaping the emission properties of these 3D relativistic flows 
  in the vicinity of rotating black holes.   
The calculations also showed the presence of limb effects 
  in 3D objects with finite optical depths.   
  
We note that gravitationally induced line resonance can occur 
  in 3D accretion onto a compact object. 
This phenomenon is not present in 2D planar objects, such as geometrically thin accretion disks,   
  where the radiation can escape from the disk surface to free space 
  without additional absorption or re-emission.

\bibliography{References}

\end{document}